\documentstyle[aps,prb,epsf,amsmath,multicol]{revtex}

\newcommand{\dir}{Figs}

\newcommand{\fig}[4]
{
    \hspace*{-1cm}
    \noindent
    \unitlength=1mm
    \begin{picture}(#2,#3)
    \put(10,0){\leavevmode \epsfxsize=#2mm \epsffile{\dir/#1}}
    \put(5,35){#4}
    \end{picture}
    \noindent
}

\begin{document}
%
%
%
\newcommand{\CCphexp}
{
\begin{figure}[h]
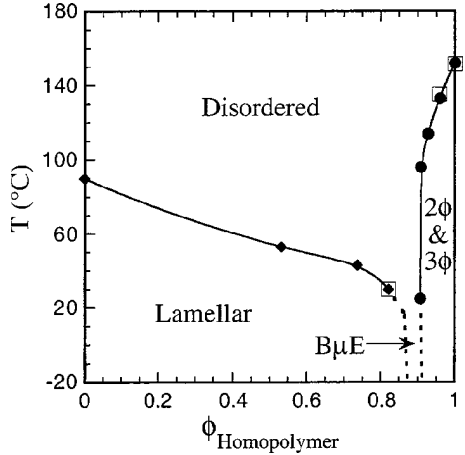

\caption{
  Experimental phase diagram \cite{hillmyer} of a PDMS-PEE/PDMS/PEE
  blend with $\alpha\approx 0.2$.
  Reproduced by permission of The Royal Society of Chemistry.
}
\label{fig:ph_exp}
\end{figure}
}
\newcommand{\CCphmf}
{
\begin{figure}
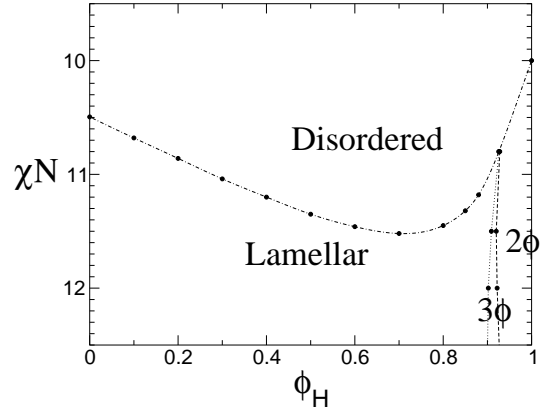

\caption{
Mean-field phase diagram for a ternary AB+A+B blend with $\alpha=0.2$,
as obtained from a Fourier space implementation of SCFT. $2\phi$ denotes a
region of two-phase coexistence between an A-rich and a B-rich phase, $3\phi$
one of three-phase coexistence between an A-rich, a B-rich, and a lamellar
phase. See text for more explanation.
}
\label{fig:ph_mf}
\end{figure}
}
\newcommand{\CChh}
{
\begin{figure}
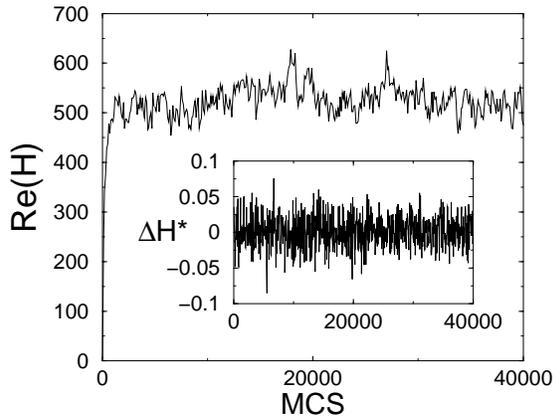

\caption{
Re(H) in a simulation of $W_-$ and $W_+$ vs. number of Monte Carlo
steps in $W_-$ for $C=50$, $\chi N=12$, $z=2$ (grand canonical ensemble),
started from a configuration equilibrated in $W_-$. Each $W_-$ step alternates
with ten steps in $\tilde\omega_+$.
Inset shows $\triangle H^*$ (difference of $\triangle H$ and a
reference value at $\tilde\omega_+$ fluctuations switched off)
in the same simulation.
}
\label{fig:hh}
\end{figure}
}
\newcommand{\CCawplus}
{
\begin{figure}
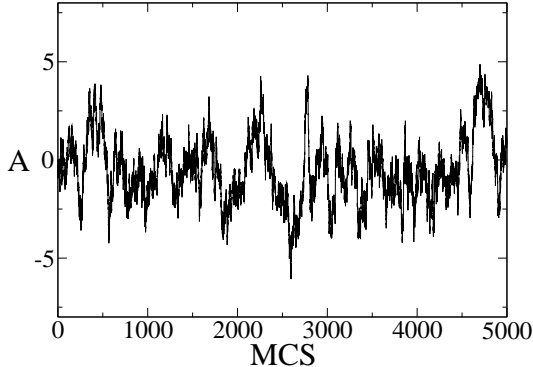

\caption{
The argument, $A$, of the cosine in the reweighting
factor in the same simulation as in Fig.~\ref{fig:hh}.
}
\label{fig:awplus}
\end{figure}
}
\newcommand{\CCwmwm}
{
\begin{figure}
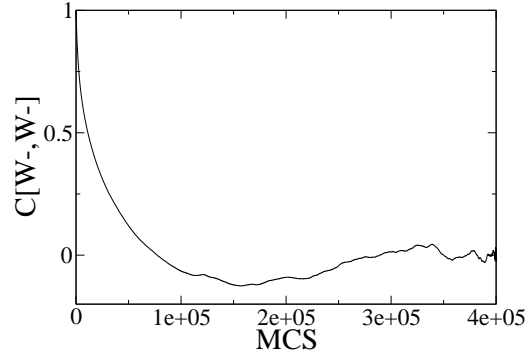

\caption{
Autocorrelation function of $W_-$ in the same simulation as in
Fig. \ref{fig:hh}. \hfill
}
\label{fig:wmwm}
\end{figure}
}
\newcommand{\CCcorrel}
{
\begin{figure}
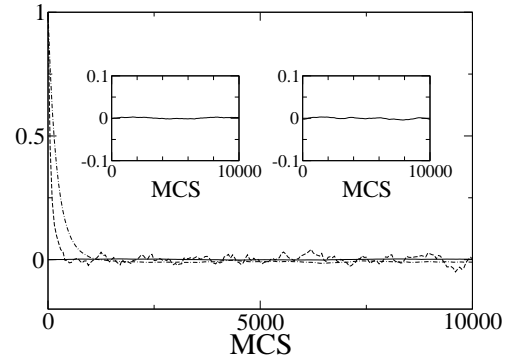

\caption{
Correlation functions $C[I,I]$ (dashed),
$C[\tilde\omega_+,\tilde\omega_+]$ (dashed-dotted), $C[I,W_-]$ (left inset),
and $C[\tilde\omega_+,W-]$ (right inset) in the same simulation as in
Fig.~\ref{fig:hh}. Note the different scale from Fig.~\ref{fig:wmwm}.
}
\label{fig:correl}
\end{figure}
}
\newcommand{\CCsnapmc}
{
\begin{figure}
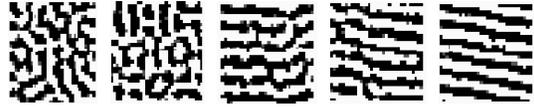

\caption{
Snapshots at $C=50$, $\phi_H=0$, and $\chi N$ = 11.1, 11.2, 11.3, 11.4,
and 11.5. White points: $0<\bar{\phi}_A<0.49$, gray points:
$0.49\le\bar{\phi}_A<0.51$, black points: $0.51\le\bar{\phi}_A\le 1$.
All runs were started from disordered configurations.
}
\label{fig:snapmc}
\end{figure}
}
\newcommand{\CClambda}
{
\begin{figure}
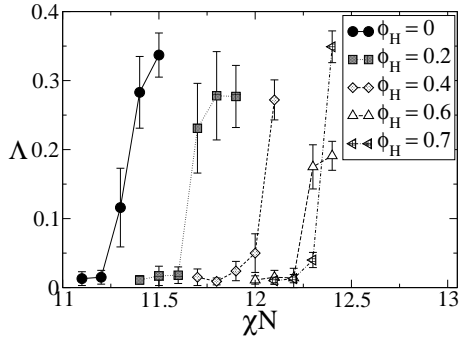

\caption{
Direction persistence as measured by the $\Lambda$ parameter
vs. $\chi N$ for different homopolymer volume fractions $\phi_H$
at the order-disorder transition in simulations on
$32 \times 32$ lattices.
For $\phi_H$ = 0 and 0.2, the discretization $\mbox{d} x=0.625$ was used,
otherwise $\mbox{d} x=0.78125$.
}
\label{fig:lambda}
\end{figure}
}
\newcommand{\CCff}
{
\begin{figure}
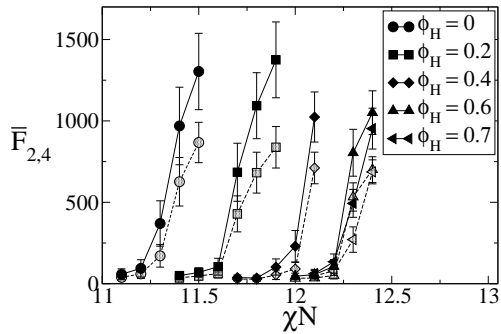

\caption{
$\bar{F}_2$ (solid) and $\bar{F}_4$ (dashed) vs. $\chi N$ for different
homopolymer volume fractions $\phi_H$ at the order-disorder
transition in the same simulations as in Fig. \ref{fig:lambda}.
}
\label{fig:f2f4}
\end{figure}
}
\newcommand{\CCfseries}
{
\begin{figure}
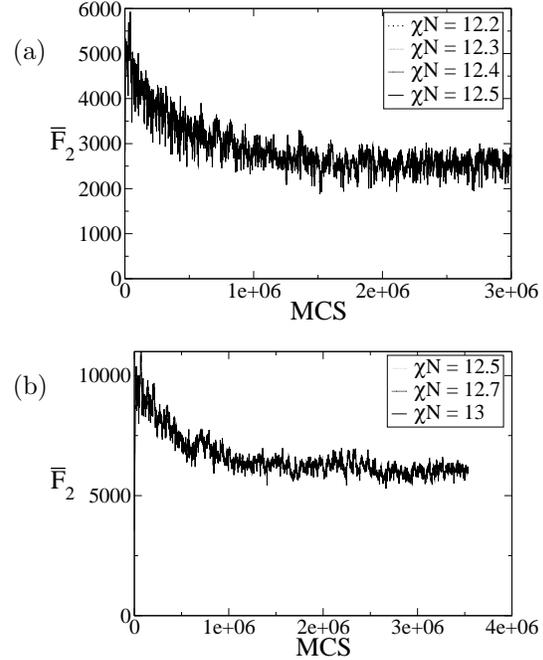

\caption{
$\bar{F}_2$ at homopolymer volume fraction $\phi_H=0.7$ (a)
and $\phi_H=0.8$ (b) for different $\chi N$ in simulations
on a $48 \times 48$ lattice that started from lamellar
configurations.
}
\label{fig:fseries}
\end{figure}
}
\newcommand{\CCconc}
{
\begin{figure}
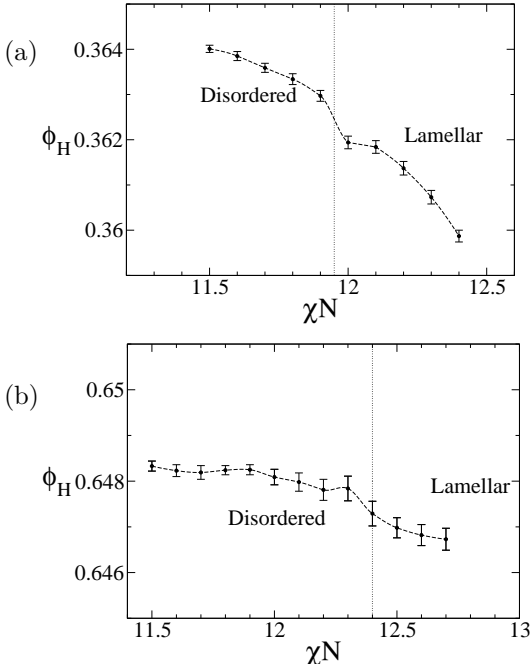

\caption{
Homopolymer concentration, $\Phi_H$, vs. $\chi N$ at
$z=1$ (a) and $z=2$ (b) across the order-disorder
transition. The dotted line indicates the
$\chi N$ coordinate of the ODT $\pm 0.1$.
}
\label{fig:conc}
\end{figure}
}
\newcommand{\CCabdemix}
{
\begin{figure}
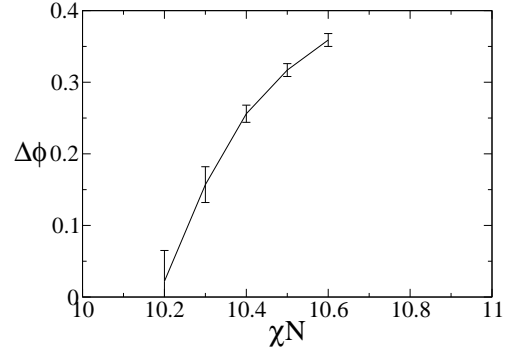

\caption{
Difference of A and B monomer densities, $\triangle \phi$, vs. $\chi
N$ in a binary system of A and B homopolymers in the grand canonical ensemble.
}
\label{fig:abdemix}
\end{figure}
}
\newcommand{\CCpc}
{
\begin{figure}
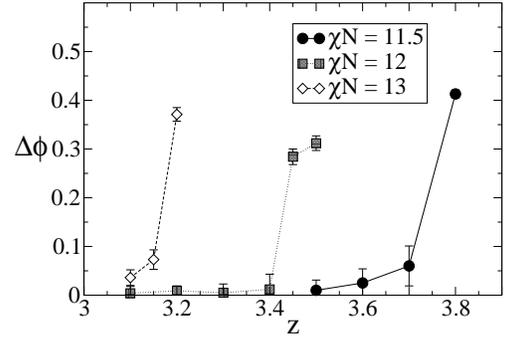

\caption{
Difference in A and B monomer densities, $\triangle \phi$,
vs. relative homopolymer fugacity, $z$, in the grand canonical ensemble.
}
\label{fig:pc}
\end{figure}
}
\newcommand{\CCvgf}
{
\begin{figure}
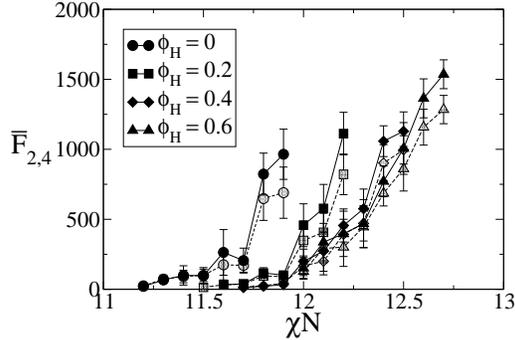

\caption{
Anisotropy parameters obtained by Complex Langevin Simulations.
The solid lines correspond to the $\bar{F}_2$ values while the dotted
lines correspond to the $\bar{F}_4$ values.
}
\label{fig:vgf}
\end{figure}
}
\newcommand{\CCsnapvg}
{
\begin{figure}
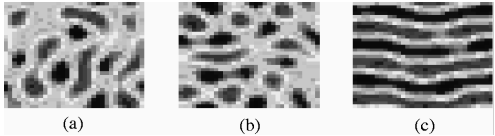

\caption{
Averaged densities across the ODT, as obtained from Complex
Langevin simulation runs. The homopolymer
fraction is fixed at $\Phi_H=0.2$. The $\chi N$ values are:
(a) $\chi N = 11.4$;
(b) $\chi N = 11.7$;
(c) $\chi N = 11.9$.
}
\label{fig:snapvg}
\end{figure}
}
\newcommand{\CCphfluc}
{
\begin{figure}
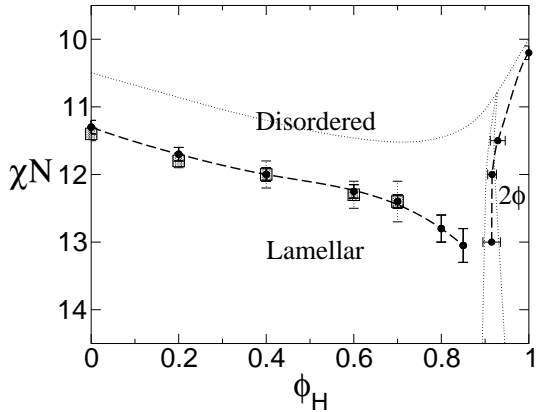

\caption{
The fluctuation-corrected phase diagram at C=50. The circles
show the results from the Monte Carlo simulations,
the squares those from the complex Langevin simulations.
The dotted lines indicate the mean-field result of
Fig. \ref{fig:ph_mf}.
}
\label{fig:ph_fluc}
\end{figure}
}
\newcommand{\CClength}
{
\begin{figure}
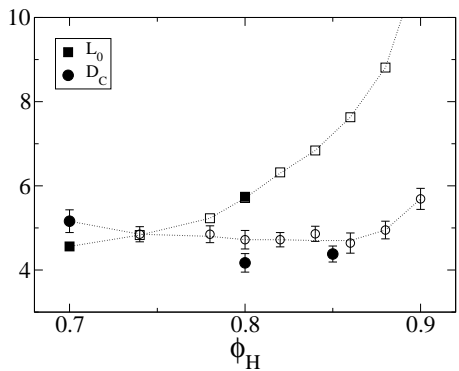

\caption{
Curvature radius $D_C$ (circles) and preferential length scale, $L_0$,
as obtained from the maximum of $S_0(q)$ (squares) at $\chi N = 12.5$
in simulations on a $48 \times 48$ lattice. Filled and empty
symbols correspond to disordered and lamellar initial conditions,
respectively. Dashed lines are guides for the eye.
}
\label{fig:length}
\end{figure}
}
%
%

\setlength{\parskip}{1.3ex}

\def\kt{k_B T}
\def\bR{ {\bf R} }
\def\br{ {\bf r}}
\newcommand{\dd}{ {\mbox{d}} }
%

\title{Fluctuation effects in ternary AB+A+B polymeric emulsions}

\author{Dominik D\"uchs$^{1}$, Venkat Ganesan$^{2}$, Glenn
        H. Fredrickson$^{3}$, Friederike Schmid$^{1}$}

\address{
         $^{1}$Fakult\"at f\"ur Physik, Universit\"at Bielefeld,
         Universit\"atsstr. 25, 33615 Bielefeld, Germany \\
         $^{2}$Department of Chemical Engineering,
         University of Texas at Austin, Austin, TX 78712 \\
         $^{3}$Departments of Chemical Engineering and Materials,
         University of California, Santa Barbara, CA 93106
        }

\maketitle

\begin{abstract}
We present a Monte Carlo approach to incorporating the effect of thermal
fluctuations in field theories of polymeric fluids.
This method is applied to a field-theoretic model of a ternary blend of AB
diblock copolymers with A and B homopolymers. We find a shift in the line
of order-disorder transitions from their mean-field values, as well as
strong signatures of the existence of a bicontinuous microemulsion phase
in the vicinity of the mean-field Lifshitz critical point.
This is in qualitative agreement with a recent series of experiments
conducted with various three-dimensional realizations of this model system.
Further, we also compare our results and the performance of the presently
proposed simulation method to that of an alternative method involving
the integration of complex Langevin dynamical equations.
\end{abstract}

\section{Introduction}

\begin{multicols}{2}
\label{introduction}
A straightforward approach to improving the properties of polymeric
materials, such as their stiffness, conductivity, toughness, is
to blend two homopolymer species (A and B) into a single melt
\cite{blends,flory,degennes,binder}. However, in most cases, weak segregation
tendencies between A and B monomers~\cite{hilde} usually cause such a
direct blend to macroscopically phase separate below a critical temperature,
or, equivalently, above a critical strength of interaction \cite{kolc}.
Such phase separation often leads to poor quality materials with
non-reproducible morphologies and weak interfaces. To circumvent this
problem, one typically introduces compatibilizers into the system,
most commonly multicomponent polymers like block, graft, or random
copolymers \cite{copolymers}. Such compatibilizers preferentially segregate
towards the interfaces between the bulk A and B domains, thereby lowering
interfacial tensions, stabilizing the formation of complex
morphologies \cite{israels,robert,mark1,balsara},
and strengthening interfaces\cite{leibler,anastasiadis,lee,pickett}.
The compatibilizers so added are usually the most expensive component
of a blend, and consequently, theoretical studies of the phase behavior
of such blends possess important ramifications in rendering such
industrial applications more economically feasible.

Here we examine the phase behavior of a ternary blend of symmetric
A and B homopolymers with an added AB block copolymer as a
compatiblizer. Specifically, we consider symmetric diblock
copolymers wherein the volume fractions
of the two different blocks in the copolymer are identical.
Moreover, the ratio of the homopolymer and copolymer lengths is
denoted as $\alpha$ and is chosen as $\alpha=0.2$. This is largely
motivated by a recent series of experiments by Bates et al.
\cite{bates1,bates2,morkved,hillmyer} on various realizations of
the ternary AB+A+B system with $\alpha \approx 0.2$. A combination
of neutron scattering, dynamical mechanical spectroscopy, and
transmission electron microscopy (TEM) was used to examine the
phase behavior. An example of an experimental phase diagram is
shown in Fig. \ref{fig:ph_exp}. It will be discussed in detail
further below.

Theoretically, one of the most established approaches to studying
such polymer blends is the self-consistent field theory (SCFT), first
introduced by Edwards \cite{edwards} and Helfand \cite{helfand} and later
used, among others, by Scheutjens and Fleer \cite{scheutjens,scheutjensbook},
Noolandi et al. \cite{noolandi}, and Matsen et al. \cite{mark1}. 
In implementing SCFT \cite{schmid}, one typically chooses (a) a model 
for the polymers and (b) a model for the interactions. Most researches 
adopt the Gaussian chain model, wherein the polymers are represented as 
continuous paths in space \cite{doi}. 
It models polymers as being perfectly flexible,
i.e., there exists no free energy penalty for bending. Instead, it
features a penalty for stretching. (In contexts where the local rigidity
or orientation effects are important, the wormlike chain model
\cite{wlc} serves as a popular alternative.) Such a molecular model
for the polymers is typically supplemented by a model for the 
interactions between unlike polymers, which is usually chosen as
an incompressibility constraint plus a generalization of the
Flory-Huggins local contact interaction term, whose strength is
governed by the product of the segregation parameter, $\chi$, and
the polymerization index, $N$.
\end{multicols}\twocolumn

\fig{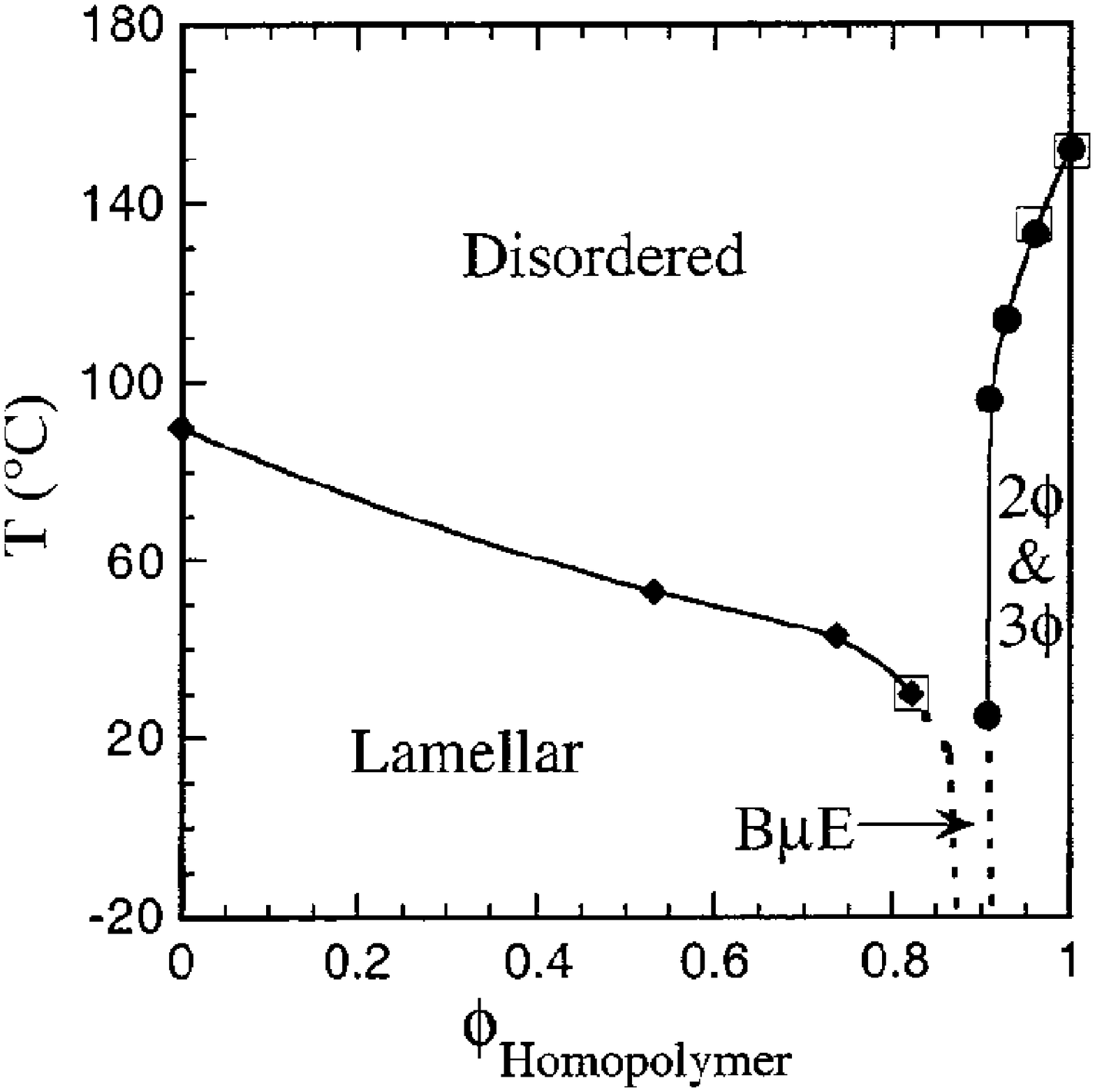}{60}{60}{}
\CCphexp

\noindent

We have implemented SCFT for the above model of polymers, and
investigated the phase behavior of the ternary A+B+AB blend. Fig.
\ref{fig:ph_mf} displays the resulting mean-field phase diagram
in the coordinates of homopolymer volume fraction $\phi_H$ and
segregation strength $\chi N$. Note that the phase diagram shown
is a slice along the ``isopleth'' where the two homopolymers are
blended in equal proportions. It can be observed that on the
copolymer-rich side of the diagram shown in Fig. \ref{fig:ph_mf},
a line of order-disorder transitions (ODT) separates the
disordered region (which occurs at low interaction strengths,
$\chi N$) from a periodically ordered (i.e., lamellar) phase at
higher $\chi N$. For pure copolymers ($\phi_H=0$), this transition
is predicted within mean-field theory to occur at $\chi N=10.495$,
and the line of ODTs, often referred to as Leibler line,
represents second-order transitions. On the homopolymer-rich side
of the diagram, we find a line of second-order transitions from a
disordered phase at low $\chi N$ to a region of two coexisting
homogeneous liquid phases at high $\chi N$. For the pure
homopolymer system ($\phi_H=1$), the upper critical consolute
point is found by mean-field Flory-Huggins theory to be at $\alpha
\chi N=2$. This line of continuous transitions is referred to as
Scott line \cite{scott}. The 
point where the Leibler and Scott
lines meet is found, again within mean-field theory, to be an
isotropic Lifshitz critical point (LP)\cite{LP}, which in the case of
$\alpha=1$ becomes a Lifshitz tricritical point (LTP)
according to Broseta and Fredrickson \cite{broseta}, and is
predicted to occur at a total homopolymer fraction of
$\phi_L = 1/(1+2\alpha^2)$ and an incompatibility of $(\chi N)_L =
2 (1+2\alpha^2)/\alpha$. 
The separation into microphases (i.e., lamellae) observed 
along the Leibler line displays a steady increase in the

\vspace*{-1cm}

\hspace*{-1cm}
\fig{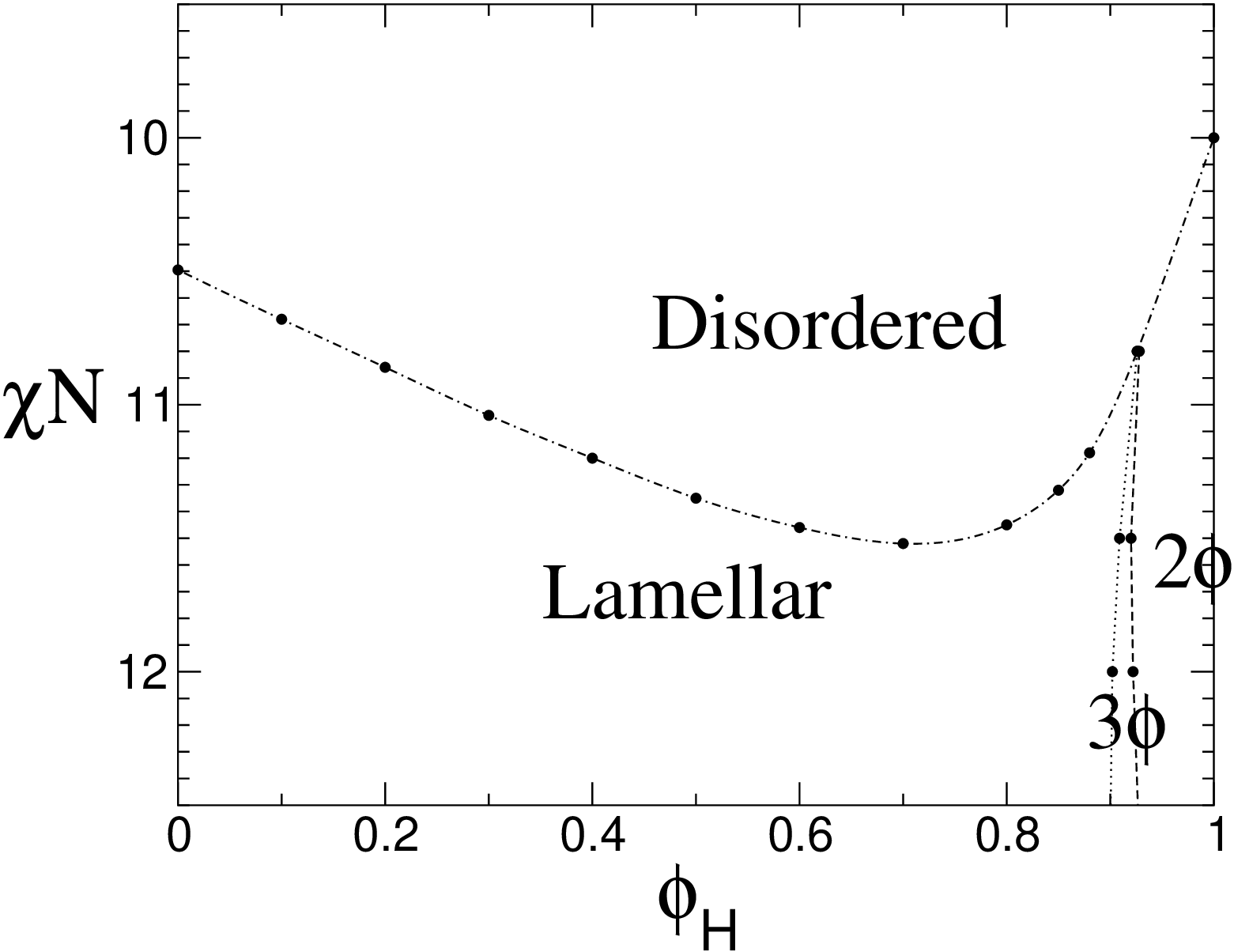}{70}{60}{}

\CCphmf

\noindent
lamellar periodicity until it finally diverges at
the LP, giving rise to macrophase separation along the Scott line.
Below the LP in the ordered regime, the system undergoes a
first-order transition from the lamellar (L) to the two-phase
(A+B) region along the axis of homopolymer concentration,
$\phi_H$. Contrary to a previous publication \cite{morkved} in
which this transition was displayed as continuous, a careful
reexamination of the mean-field theory reveals that it is indeed
first-order with a three-phase (L+A+B) coexistence region reaching
all the way up to the LP for any $\alpha$. This is further
corroborated by recent results of Naughton and Matsen\cite{naughton}.

The mean-field phase diagram displayed in Fig. \ref{fig:ph_mf} differs
substantially from the experimental phase diagram, shown in
Fig. \ref{fig:ph_exp}. In the experiments, the LP was destroyed
by fluctuations, giving rise to a bicontinuous microemulsion phase in its
vicinity. At lower
temperature $T$, the L and A+B phases were conjectured to be
separated by a narrow channel of this microemulsion phase
stretching down from the Lifshitz region, while the Leibler and
Scott lines were shifted from their mean-field locations to lower
temperatures (i.e., higher $\chi N$, which is proportional to
$1/T$). Experimental results
are consistent with the picture that as the lamellar periodicity
increases along the $\phi_H$ axis, the persistence length of
composition fluctuations along the microdomain boundaries
decreases until at some point the two become comparable in size
and the lamellae begin to rupture forming a bicontinuous
structure.

The experimental results discussed above, while in contradiction to the
mean-field SCFT calculations, are not entirely surprising when viewed
in the context of theoretical studies pertaining to the effect of
fluctuations on the mean-field phase diagrams. Indeed, even in the context
of the pure diblock copolymer (i.e.~$\phi_H=0$), Leibler \cite{leibler2}
speculated that the order-disorder transition (predicted to be second
order within the mean-field theory) would become a weakly first-order
transition when the effects of fluctuations are accounted for. Such a
reasoning was confirmed quantitatively by Fredrickson and
Helfand\cite{fredrickson}, who extended an earlier analysis
by Brazovskii \cite{brazovskii} to show that fluctuations change
the order of the transition and also shift the transition
to higher incompatibilities, $\chi N$. Further, they quantified such
shifts in terms of a parameter that depends on the
 polymer density (or, equivalently in the context of a polymer
melt, their molecular mass $N$). This parameter, denoted generically
as $C$ in this text, acts as a Ginzburg parameter \cite{ginzburg}, and in
the limit $C\rightarrow\infty$, the mean-field solution is recovered.

The present study is motivated by the hypothesis that the fluctuation effects
not accounted for within SCFT are responsible for the discrepancies noted
between the experimental results and the theoretical predictions for
the ternary-blend phase behavior. As a partial confirmation
of our hypothesis, the experimental phase
diagrams (Fig. \ref{fig:ph_exp}) of Bates and co-workers deviate most
clearly from the mean-field diagram for the PEE/PDMS/PEE-PDMS system,
in situations where the molecular weights were much lower than in
the other systems (i.e., where the $C$ parameter is small). This article
presents a theoretical examination of the effect of thermal
fluctuations upon the phase behavior of ternary blends. Our
primary goal is to understand and quantify the shift in both the
Leibler and Scott lines and to examine the formation of the microemulsion
phases in such systems. Bicontinuous structures such as the
ones observed in the above experiments are particularly
interesting from an application point of view in that they are
able to impart many useful properties on polymer alloys. For
example, if one blend component is conducting or gas-permeable,
these properties will be passed on to the entire alloy. Also,
mechanical properties such as the strain at break and the
toughness index have been observed to exhibit maxima with
co-continuous structures \cite{bicon}.

Previous simulation studies of fluctuation effects in ternary
diblock copolymer / homopolymer blends 
\cite{marcus,andreas,kielhorn,liang,poncela},
have mostly focused on the $\alpha=1$ case. Within a lattice
polymer model, M\"uller and Schick \cite{marcus} have investigated 
the phase diagram of a corresponding symmetrical A+B+AB blend. 
According to their results, the disordered phase extends to higher 
$\chi N$ than predicted by the mean field theory, and the
LTP is replaced by a regular tricritical point. Due to finite 
size effects, the transition to the lamellar phase could not be 
localized quantitatively.

Particle-based models have the disadvantage that one is restricted 
to studying relatively short polymers. This motivates the use of
field theories in simulations. Kielhorn and Muthukumar \cite{kielhorn} 
have studied the phase separation dynamics in the vicinity of the LTP
within a Ginzburg-Landau free energy functional, which they derived 
from the Edwards model \cite{edwards} within the random phase 
approximation (RPA), in the spirit of the theories for pure diblock 
copolymer systems mentioned earlier \cite{leibler2,fredrickson}. 
Naively, one should expect that such a description is adequate in 
the weak segregation limit, where the correlation length for 
composition fluctuations is large. However, Kudlay and Stepanow
\cite{kudlay} have recently questioned this assumption.
They pointed out that the fluctuations in RPA-derived Hamiltonians 
have significant short wavelength contributions which cannot be 
eliminated by renormalization and may lead to unphysical predictions.

In this work, we propose an alternative field-theoretic approach which 
rests directly on the Edwards model, without resorting to the additional
RPA approximation. The partition function is evaluated with a Monte 
Carlo algorithm. A related algorithm, the complex Langevin method,
has been presented recently by two of us \cite{venkat,glenn,glennreview}. 
Here we study two-dimensional ternary blends at $\alpha = 0.2$ with
both methods and compare the results.

The remainder of this article is organized as follows. In section
\ref{model}, we outline the field-theoretic formulation of our model.
The simulation method is presented in section \ref{simulations}.
In section \ref{results},
we present and discuss our results before we conclude in section
\ref{summary} with a summary and an outlook on future work.

\section{The field-theoretic model}
\label{model}
In this section, we present a field-theoretic model for the system under
consideration. To this effect, we consider a mixture of $n_A$ homopolymers
of type A, $n_B$ homopolymers of type B, and $n_{AB}$ symmetric block
copolymers in a volume $V$. The polymerization
index of the copolymer is denoted $N$, and the corresponding
quantities for the homopolymers are denoted $N_A = N_B = \alpha N$.
All chains are assumed to be monodisperse. We consider the case of
symmetric copolymers, where the fraction of A monomers in the
copolymer is $f = 1/2$. We restrict our attention to the
concentration isopleth, where the homopolymers have equal volume
fractions $\phi_{HA}=\phi_{HB}=\phi_{H}/2$. The monomeric volumes
of both A and B segments are assumed to be identically equal to
$1/\rho_0$. In this notation, the incompressibility constraint
requires
\begin{equation}
n_A=n_B=\frac{\phi_H \rho_0 V}{2 \alpha N}\mbox{;} \:
n_{AB}=(1-\phi_H)\frac{\rho_0 V}{N},
\end{equation}
where $V$ denotes the volume of the system. In the Gaussian chain model,
the segments are assumed to be perfectly flexible and are assigned a common
statistical segment length $b$. The polymers are represented as continuous
space curves, ${\bf R}_j^i (s)$,
where $i$ denotes the polymer species (A, B, or AB), $j$ the different
polymers of a component, and $s$ represents the chain length variable
measured along the chain backbone such that $0\le s\le \nu_j$ with
$\nu_{AB}=1$ and $\nu_A=\nu_B=\alpha$. Conformations of non-interacting
polymers are given a Gaussian statistical weight, $\exp (-H_0 )$, with a
harmonic stretching energy given by (units of $k_B T$),
\begin{equation}
\label{eq4}
H_0 [ \bR ] = \frac{3}{2Nb^2} \sum_{i} \sum_{j=1}^{n_j} \int_0^{\nu_i}\
\dd s \;
{\bigg |} \frac{{d {\bR_j^i (s)}}}{ {ds}}{\bigg |}^2  .
\end{equation}
where $b_A = b_B = b_{AB} \equiv b$ is the statistical segment
length. Interactions between A and B segments are modeled as local
contact interactions by a pseudopotential with a Flory-Huggins
quadratic form in the microscopic volume fractions,
\begin{equation}
\label{eq5a}
H_I [ \bR ] = \chi \rho_0 \int \dd {\br} \,
\hat\phi_A(\br) \hat\phi_B(\br)
\end{equation}
which with the substitutions
$\hat{m}=\hat\phi_B-\hat\phi_A$, $\hat\phi=\hat\phi_A+\hat\phi_B$
transforms into
\begin{equation}
\label{eq5}
H_I [ \bR ] = \frac{\chi \rho_0}{4} \int \dd {\br} \, (\hat{\phi}^2
- \hat{m}^2).
\end{equation}
$\chi$ denotes the Flory-Huggins parameter, and the microscopic density
operators $\hat{\phi}_A {\text{ and }} \hat{\phi}_B$ are defined by
\begin{eqnarray}
\hat{\phi}_A (\br) & = & \frac{N}{\rho_0} \sum_{j=1}^{n_{AB}}
\int_0^{f} \dd s \; \delta (\br - \bR_j^{AB} (s))
\nonumber \\
&& + \: \frac{N}{\rho_0}
\sum_{j=1}^{n_{A}} \int_0^{\alpha} \dd s \;
\delta (\br - \bR_j^A (s)) \\
\hat{\phi}_B (\br) & = & \frac{N}{\rho_0} \sum_{j=1}^{n_{AB}}
\int_{f}^{1} \mbox{d} s \; \delta (\br - \bR_j^{AB} (s))
\nonumber \\
&& + \: \frac{N}{\rho_0} \sum_{j=1}^{n_{B}} \int_0^{\alpha} \dd s
\; \delta (\br - \bR_j^B (s)).
\end{eqnarray}
Furthermore, the incompressibility constraint is implemented by the insertion 
of a Dirac delta functional~\cite{footnote}, 
and we obtain the canonical partition function of the blend,
${\cal Z}_C$:
\begin{equation}
\label{eq7}
{\cal Z}_C \propto \int  \prod_{i} \prod_{j}
D \bR_j^i (s) \;  \delta (\hat{\phi} - 1) \; \exp(- H_0- H_I ),
\end{equation}
where the $\int D \bR$ denote functional integrations over chain
conformations. Representing the delta functional in exponential form and
employing a Hubbard-Stratonovich transformation on the $\hat{m}^2$ term in
(\ref{eq5}), we obtain
\begin{equation}
\label{eq8}
{\cal Z}_C \propto \int_{\infty} D W_-
\int_{i\infty} D W_+ \exp [ - H_C (W_-, W_+)]
\end{equation}
with
\begin{eqnarray}
\label{eq:hc}
\lefteqn{ H_C (W_-, W_+) =
C \Big[\frac{1}{\chi N} \int \dd {\br} \, W_-^2
- \int \dd {\br} \, W_+}
\nonumber \\ &&
- V(1 \!-\!\phi_H) \ln Q_{AB}
- \frac{V\phi_H}{2\alpha} (\ln Q_A + \ln Q_B) \Big]
\end{eqnarray}
\begin{equation}
C=\frac{\rho_0}{N} R^d_{g_0}.
\end{equation}
Here and throughout this article, all lengths are expressed in units of the
unperturbed radius of gyration, $R_{g_0}=b(N/(2d))^{1/2}$, where $d$ is the
space dimension. The parameter $C$ in the above equations, which occurs
as a global prefactor to $H_C$, acts as a Ginzburg parameter such
that in the limit
$C\rightarrow\infty$ the partition function (\ref{eq8}) is
dominated by its saddle point and the mean-field solution becomes
exact. In (\ref{eq:hc}), $i \equiv \sqrt{-1}$, and $Q_A, Q_B,
{\text{ and }} Q_{AB}$ denote respectively the single-chain
partition functions for the A, B, and AB chains, in the potential
fields $W_- (\br) {\text{ and }}W_+ (\br)$. Note that $W_-$ is
conjugate to the difference in A and B densities, $\hat m$, and
$W_+$ to the total density, $\hat\phi$. Moreover, $W_-$ is real,
whereas $W_+$ is imaginary, thereby rendering $H_C$ complex.

The single-chain partition functions can be expressed in terms of the
Feynman-Kac formulae \cite{helfand} as:
\begin{equation}
\label{eq13}
Q_i = \int \dd \br \; q_i (\br ,\nu_i) ,
\end{equation}
where the propagators $q_i$ satisfy the diffusion
equations
\begin{equation}
{\frac{\partial}{\partial s}} q_i (\br ,s) = \triangle q_i (\br ,s) - q_i
U_i;\quad q_i(\br, 0)=1
\label{diff}
\end{equation}
\begin{equation}
\label{eq11}
\begin{split}
U_A & = W_+ - W_-, \hspace{0.5cm} 0 \le s \le \alpha\\
U_B & = W_+ + W_-, \hspace{0.5cm} 0 \le s \le \alpha \\
U_{AB} & =
\begin{cases}
U_A, & 0 \le s < f \\
U_B, & f \le s \le 1
\end{cases}
\end{split}
\end{equation}
An analogous diffusion equation applies to the conjugate
propagators, $q^\dagger_i$, which propagate from the opposite end
of a polymer. Due to their symmetry, the propagators of the
homopolymers are their own conjugates. Hence, we can calculate
density
operators, $\bar\phi_A$ and $\bar\phi_B$, from
\begin{eqnarray}
{\bar{\phi}}_A (\br) & = & \frac{V(1-\phi_H)}{Q_{AB}}
\int_0^f \!\!\! \dd s \;
q_{AB}(\br ,s) \: q^\dagger_{AB} ( \br ,1\!-\!s)
\nonumber\\
&+& \: \frac{V\phi_H}{2 \alpha Q_{A}} \int_0^\alpha \!\!\! \dd s
\; q_A (\br ,s) \: q_{A} ( \br ,\alpha-s),
\label{eq:phi}
\\
{\bar{\phi}}_B (\br) &=&  \frac{V(1 - \phi_H)}{Q_{AB}} \int_f^1
\!\!\! \dd s \;
q_{AB} (\br,s) \: q^\dagger_{AB} ( \br ,1\!-\!s)
\nonumber\\
&+& \frac{V\phi_H}{2 \alpha Q_{B}} \int_0^\alpha \!\!\! \dd s
\; q_B (\br ,s) \: q_{B} ( \br ,\alpha-s).
\label{barphi}
\end{eqnarray}
These d shown on these snapshots are not to be confused
with real densityensity operators depend on $W_-$ and $W_+$ and are in
general complex. However, their ensemble averages yield real
densities which correspond to the experimentally measurable quantities:
$\phi_{A,B} = \langle\bar\phi_{A,B}\rangle (= \langle\hat\phi_{A,B}\rangle)$.

The above derivation assumed a canonical ensemble for the blend, wherein the
average compositions of the different compositions are fixed. For future
reference, we also list a similar set of formulae obtained in the
grand-canonical ensemble:
\begin{eqnarray}
H_{GC} (W_-, W_+) &=& C \Big[\frac{1}{\chi N} \int \dd {\br} \, W_-^2
- \int \dd {\br} \, W_+
\nonumber \\ &&
\label{eq:hgc}
- Q_{AB} - z Q_A - z Q_B \Big],
\end{eqnarray}
where $z=\exp((\triangle\mu/k_B T)$,and
$\triangle\mu\equiv\mu_A-\mu_{AB}=\mu_B-\mu_{AB}$ is
the difference in the chemical potentials of homopolymers and copolymers. 
The density
operators are given as:
\begin{eqnarray}
{\bar{\phi}}_A (\br)& := & \int_0^f \dd s \;
q_{AB}(\br ,s) q^\dagger_{AB} ( \br ,1-s)
\nonumber \\ &&
+ z \int_0^\alpha \dd s \; q_A (\br ,s) q_{A} ( \br ,\alpha-s),
\label{eq:phigc}
\\
{\bar{\phi}}_B (\br)& := & \int_f^1 \dd s \;
q_{AB} (\br ,s) q^\dagger_{AB} ( \br ,1-s)
\nonumber \\ &&
+ z \int_0^\alpha \dd s \; q_B (\br ,s) q_{B} ( \br ,\alpha-s).
\end{eqnarray}

\section{Field-theoretic simulations}
\label{simulations}

\subsection{General considerations}
Field-theoretic models like the one introduced above have been used in many
earlier researches, and the resulting functional integrals of the partition
function have been analyzed using various approximate methods. In
self-consistent mean-field theory (SCFT)\cite{mark1,schmid,drolet}, the
integrand of (\ref{eq8}) is approximated by saddle points ($W_+^*, W_-^*$)
such that
\begin{equation}
\frac{\delta H}{\delta W_+}{\bigg |}_{W_+^*, W_-^*} = 0, \qquad
\frac{\delta H}{\delta W_-}{\bigg |}_{W_+^*, W_-^*} = 0.
\label{eq15}
\end{equation}
A homogeneous saddle point then corresponds to the disordered
phase, whereas every ordered phase has a unique inhomogeneous
saddle point. By examining (\ref{eq8}), it is further apparent that $C$
plays the role of a Ginzburg parameter in the sense that the mean-field
approximation, i.e.~$Z \simeq \exp( - H [W_+^*, W_-^*])$,  becomes exact
in the formal limit of $C \rightarrow \infty$, corresponding to
$N \rightarrow \infty$ for $d > 2$. Implementations of
SCFT \cite{mark1} can thus be viewed as numerical strategies for
computing the various saddle points. However, thermal
fluctuation effects in the vicinity of phase boundaries invalidate
the mean-field results in these regions, and strategies for
evaluating (\ref{eq8}) must be found. Analytical methods are quite
limited, especially for systems dominated by inhomogeneous saddle
points, and often involve weak segregation expansions \cite{brazovskii}.
In contrast, earlier publications of two of us
\cite{venkat,glenn,glennreview} have detailed a new approach termed
{\em field-theoretic simulations}, which is a methodology for
numerically sampling functional integrals in field-theoretic models of
polymer solutions and melts. The underlying idea is to numerically
sample $Z$ (independently of the type of ensemble employed) by generating
a Markov chain of $W_-, W_+$ states with a stationary distribution
proportional to $\exp ( - H [W_-, W_+])$. Averages of physical quantities
${\bar{f}} [W_-, W_+]$ can then be approximated by ``time'' averages
using the states of the Markov chain,
\begin{equation}
\langle {\bar{f}} [W_-, W_+] \rangle \approx {1}/{M} \sum_{j = 1}^{M}
{\bar{f}} [W_{-,j}, W_{+,j} ],
\end{equation}
where $j$ labels the states of the Markov chain. True equality of ensemble
and time averages is established as usual for ergodic systems in the limit
$M \rightarrow \infty$. Such a numerical method is non-perturbative in
nature, allowing for a more complete account of fluctuation effects.

In this article, we undertake field-theoretic simulations to
address the effect of fluctuations upon the mean-field phase
diagram. Note that to numerically generate the Markov chain of
$W_-, W_+$ states, one needs to account for the fact that $H [W_-,
W_+]$ ((\ref{eq:hc}) and (\ref{eq:hgc})) in general possesses both
real and imaginary parts, implying a non-positive-definite
statistical weight $\exp (-H)$. In our  previous researches we
implemented a general ``complex Langevin dynamics'' technique that
had been proposed earlier to deal with such non-positive-definite
statistical weights \cite{parisi,gausterer}. In view of the fact
that this method has been discussed in detail in our previous
publications, we restrict our discussion to a brief outline of the
technique which is presented in the Appendix. It is to be noted
that outstanding issues still remain as regards the theoretical
foundations of this technique \cite{parisi}. While in our earlier
studies we encountered no problems with the convergence or the
uniqueness of the solutions, in the present study this method did
encounter numerical difficulties, especially in the regions close
to the Lifshitz point.
These difficulties are associated with strong fluctuations and
complex phase oscillations and will be described more fully in a
subsequent section.
%

In an effort to probe and overcome the
difficulties encountered in preliminary complex Langevin
simulations of the present ternary blend and quantify the effect
of fluctuations on the
phase diagram of this system,
we present an alternative, albeit equally rigorous
technique for implementing field-theoretic simulations, which we
refer to as the {\em field-theoretic Monte Carlo method}. In the
discussion of our results, we compare explicitly the results
obtained by the latter method with that obtained through the
complex Langevin method
and
find good agreement in the range of their validities. It is
pertinent to also note that there also exist real Langevin methods
in the literature that focus specifically on the $W_-$
fluctuations \cite{reister}.

\subsection{The field-theoretic Monte Carlo method}
Our approach to sampling (\ref{eq8}) is based on the Monte Carlo
method. We recall that $W_-$ is real and $W_+$ imaginary, and that
the fluctuations in $W_+$ are conjugate to the total density.
Although the incompressibility constraint implemented in our model
effectively only constrains
$\phi\equiv\phi_A+\phi_B=\langle\bar\phi_A+\bar\phi_B\rangle$,
one might guess that the effect of the $W_+$ fluctuations that
impose this constraint is rather small compared to the
composition fluctuations governed by $W_-$. Therefore, we shall
tackle $W_-$ and $W_+$ in two distinct steps, after which we will
be able to argue that the effect of the $W_-$ fluctuations is
indeed the dominant one.

To simulate the fluctuations in $W_-$, we pick a starting density 
configuration and use the SCFT mean-field equations \cite{schmid} 
to calculate a seed $W_-$ and $W_+$. 
Next, we proceed according to the following scheme:

(a) Make a global move in $W_-$, i.e., tentatively add a random
number $\in[-1,+1]$ times a step width parameter to every $W_-(\br)$, and
denote by $W_-^0$ the old $W_-$.

(b) Calculate a self-consistent $\bar W_+$ that is a partial
saddle point with the tentative new $W_-$ fixed; to do this, we
need to iteratively solve the equation
\begin{equation}\label{}
  \bar{\phi}_A (\br ; [\bar W_+ ,W_- ]) + \bar{\phi}_B (\br ; [\bar W_+ ,W_-
  ]) = 1
\end{equation}
for $\bar W_+ (\br )$. The initial guess for $\bar W_+$ in this iteration,
$\bar W_+^0$, is taken to be the old $\bar W_+$ from the last Monte Carlo
cycle. The iteration, which employs a two-step Anderson mixing scheme
\cite{eyert}, is terminated once $\triangle H\equiv H[W_-,\bar
W_+]-H[W_-^0,\bar W_+^0]$ is determined to within 0.001, or a few
parts in $10^4$. Every step in this iteration requires the
solution of the diffusion equation (\ref{diff}) on the entire
simulation lattice, and this is where practically all the
computing time is spent. The number of iterations to find $\bar
W_+$ in most cases is roughly around 10.

(c) Accept or reject the move according to a standard Metropolis
criterion. In other words, we always accept the move if the resulting
difference in $H$, i.e.,~$\triangle H$, is negative, and
otherwise, we accept with a probability $\exp(-\triangle H)$.

(d) Go back to (a) unless the maximum number of Monte Carlo steps has
been reached.

Next, we turn to discuss the sampling of $W_+$ fluctuations.
As a result from SCFT, we know that the saddle point $W_+^*$ of $W_+$ as well
as the partial saddle point $\bar W_+[W_-]$ are purely real, despite the fact
that the integration path of $W_+$ in (\ref{eq8}) is along the imaginary
axis. Neglecting surface terms and because we do not cross any singularities,
we can therefore deform the integration path and represent $W_+$ as
\begin{equation}
W_+ (\br) = \bar W_+[W_-] (\br) + i \tilde\omega_+ (\br),
\end{equation}
where $\tilde\omega_+ (\br )$ is real. Furthermore, we split the
argument of the integral in the partition function into a real
part and a complex reweighting factor, leading to the following
expression for the canonical ensemble:
\begin{equation}
Z_C \propto \int {\cal D}W_- \int {\cal D}\tilde\omega_+
\exp\Big(-H_C^R\Big) \cdot I_C,
\end{equation}
\begin{eqnarray}
H_C^R &=& C\Big[\frac{1}{\chi N}\int \dd \br W_-^2 - \int \dd \br\bar W_+
\nonumber\\ &&
\label{eq:hcr}
- \: \sum_i \Big( \frac{V_i}{\nu_i} \mbox{ Re} (\ln
Q_i[W_-,\tilde\omega_+])\Big)\Big],
\end{eqnarray}
\begin{eqnarray}
I_C &=& \exp\Big(C\Big[i \int \dd \br\tilde\omega_+
\nonumber\\ &&
+ \: i \sum_i\frac{V_i}{\nu_i}
\mbox{Im} \big( \ln Q_i[W_-,\tilde\omega_+]\big)\Big]\Big).
\end{eqnarray}
where $V_i$ is the volume occupied by species $i$. In the grand
canonical ensemble the corresponding expression is:
\begin{eqnarray}
\label{Z_GC}
Z_{GC} &\propto& \int \! {\cal D}W_- \! \int \! {\cal D}\tilde\omega_+
\exp\left(-H_{GC}^R\right) \cdot I_{GC},\!\!
\\
H_{GC}^R &=& C\Big[\frac{1}{\chi N}\int \dd \br W_-^2 - \int \dd \br\bar W_+
\nonumber \\ &&
\label{eq:hgcr}
- \: \sum_i z_i \mbox{ Re} \big(Q_i[W_-,\tilde\omega_+])\big)
\Big],
\\
I_{GC} &=& \exp\Big(C\Big[i \int \dd \br \tilde\omega_+
\nonumber \\ &&
\label{I_GC}
+ \: i \sum_i z_i \mbox{ Im} \big( Q_i[W_-,\tilde\omega_+]\big)\Big]\Big),
\end{eqnarray}
where $z_{AB}=1$ and $z_A=z_B=z$.

We can thus in principle simulate both $W_-$ and $W_+$ fluctuations, e.g., by
alternating a move in $W_-$ with a number of moves in $\tilde\omega_+$ 
and replacing $\triangle H_{C,GC}$ with $\triangle H^R_{C,GC}$ in the 
Metropolis criterion. Each configuration then needs to be weighted with 
the factor $I_{C,GC}$
for the purpose of computing averages. We note that the
propagators and the field term in the diffusion equation
(\ref{diff}) are now complex, but the averages of all physical
quantities will approach real values in equilibrium.

At this point, a remark concerning our numerical methods is due: 
the single most time-consuming part in doing our simulations is the 
solution of the diffusion equation (\ref{diff}). Initially, we
used either the Crank-Nicholson (CN) or DuFort-Frankel (DF) finite
differencing schemes \cite{recipes,fdm} to effectuate this task.
As it turned out, however, a pseudo-spectral method that had been
used successfully in the past to solve the mathematically very
similar Schr\"odinger equation and that had recently been applied
to SCFT \cite{feit,rasmussen} was far better suited for our needs.
Typically, it allowed for an order of magnitude higher values of
contour step $\dd s$ than either CN or DF to achieve the same
accuracy. Implementing the scheme with highly optimized Fast
Fourier Transform routines like the publicly available FFTW
library \cite{fftw}, the computing time per contour step is only
roughly double that needed for DF and half that needed for CN on a
single processor. We therefore regard it as the method of choice
in the context of polymer field theory. For the larger system
sizes to be discussed in the next section, a parallel version of
the code was used.

\subsection{Analysis of $W_+$ fluctuations}
Evidently, the $W_+$ fluctuations exert their influence on the simulation
in two ways: (a) through the complex reweighting factor $I_{C}$ or $I_{GC}$
and (b) through the difference of $\triangle H^R$ from $\triangle H$.

We consider first $I_{GC}$ and to this end split the complex diffusion equation
into two real equations for the real and imaginary parts of the propagator for
polymers of type $i$, denoted $q_i^R$ and $q_i^I$, respectively.
Note the symmetry properties of this coupled set of equations with respect to a
sign change in $\tilde\omega_+$: for $W_+ = \bar W_+ \pm i\tilde\omega_+$ we
obtain
\begin{equation}
\begin{array}{*{10}{lllllllll}}
&\frac{\partial}{\partial s} q_i^R &=& &\triangle q_i^R& - &U_i q_i^R& \pm
&\tilde\omega_+ q_i^I&, \\
\pm&\frac{\partial}{\partial s} q_i^I &=& \pm&\triangle q_i^I& \mp &U_i q_i^I& -
&\tilde\omega_+ q_i^R& .
\end{array}
\label{complexdiff}
\end{equation}
We see that $q_i^R$ remains unchanged whereas $q_i^I$ goes to $-q_i^I$. Thus, for
the corresponding single-chain partition function $Q_i=Q_i^R+iQ_i^I$, $Q_i^R$
is symmetric in $\tilde\omega_+$, and $Q_i^I$ antisymmetric. Furthermore, the
exponent of the reweighting factor $I_{GC}$ is antisymmetric in
$\tilde\omega_+$ (cf. (\ref{I_GC})). Since in the partition function
(\ref{Z_GC}) we integrate
over all possible configurations of $\tilde\omega_+$, the imaginary parts of
$I_{GC}$ cancel out for each antisymmetric pair, and consequently, we
effectively only need to apply the real part of $I_{GC}$:
\begin{equation}
I_{GC}^R = \cos\left( C\left[\int \dd \br \tilde\omega_+ + \sum_i z_i
Q^I_i[W_-,\tilde\omega_+]\right]\right).
\end{equation}

These symmetries reveal that in a functional expansion of $Q_i$ in
$\tilde\omega_+$ around the partial saddle point, $\bar Q_i[W_-]$, the
even-order
terms coincide with the expansion of $Q_i^R$, and the odd-order terms with that
of $Q_i^I$. Moreover,
\begin{equation}
\frac{\delta Q_i}{\delta W_+} \equiv \frac{\delta Q_i^I}{\delta
\tilde\omega_+},
\end{equation}
and
\begin{equation}
\sum_i z_i \frac{\delta Q_i^I}{\delta \tilde\omega_+} = \sum_i z_i
\frac{\delta Q_i}{\delta W_+} = -\langle\hat\rho\rangle = -1
\end{equation}
by virtue of the incompressibility constraint. Hence,
\begin{equation}
\sum_i z_i Q_i^I = - \int \dd \br \tilde\omega_+  + {\cal O}(\tilde\omega_+^3),
\end{equation}
and finally
\begin{eqnarray}
\label{O6}
I_{GC} &=& \cos\left({\cal O}(\tilde\omega_+^3)\right) = 1 - {\cal
O}(\tilde\omega_+^6), \\
\label{O2}
Q_i^R &=& \bar Q_i[W_-] + {\cal O}(\tilde\omega_+^2).
\end{eqnarray}

Based on these analytical results, we should expect $I_{GC}$ to
deviate only very slightly from unity so that it can be neglected
in most practical cases. Hence, the impact of the $W_+$
fluctuations would materialize solely as a (real) ${\cal
O}(\tilde\omega_+^2)$ contribution to $H_{GC}$. The above
derivation of (\ref{O6}) and (\ref{O2}) was for the grand
canonical ensemble but holds true if one is to derive analogous
relations in the canonical ensemble, as well, because the leading
term of $\ln Q_i$ is linear in
$\tilde\omega_+$, as in the grand canonical ensemble. It should
however be noted that in the canonical ensemble, the gauge
invariance of the
theory to a constant shift of the $\tilde\omega_+$ field
allows the introduction of a spurious phase
factor $e^{i\psi}$ that preserves $H_C$ (and hence $|Q_i|$ and
$H^R$). This $\psi$ field will freely move in canonical
simulations unless constrained, e.g., by demanding that $\int \dd
\br\tilde\omega_+(\br)=0$. In contrast, in the grand canonical
ensemble, $Q^R$ and hence $H^R$ are to lowest approximation
quadratic functions around $\tilde\omega_+ = 0$. As a result the
shape of the $H^R(W_+)$ energy landscape is much steeper in the
grand canonical ensemble than in the canonical, and no constraint
is needed.

While the above considerations can only give us an indication of the
insignificance of the $W_+$ fluctuations, we have also done test
simulations to get more conclusive estimates. We therefore choose the grand
canonical ensemble for testing our simulation technique of $W_+$
fluctuations. Before starting a run at $z = 2$, $\chi N=12$, $\alpha=0.2$,
and $C=50$, we fully equilibrated a simulation of $W_-$ fluctuations
only with the same parameters, which started from a disordered
configuration and ran a million Monte Carlo steps. This set of
parameters is very close to the (fluctuation-corrected)
order-disorder transition in the
phase diagram, on the disordered side. The average homopolymer
fraction, $\phi_H$, is  $0.65$. On a 32$ \times $32 lattice with periodic
boundary conditions, we used $\dd x=0.78125$. We then took this
equilibrated configuration and switched on the $W_+$ fluctuations,
again making ten (global) moves in $W_+$ after each move in $W_-$
so as to roughly spend equal amounts of computing time for the two
types of moves. Note in Fig. \ref{fig:hh} how $H_{GC}$ quickly
increases and then plateaus off, indicating that the $W_+$
fluctuations have become saturated.

For the same run, we recorded the quantity
$\triangle H^*= \triangle H^R - \triangle H$
(cf. Eqs. (\ref{eq:hcr}) and (\ref{eq:hgcr})),
which is displayed in the inset of Fig. \ref{fig:hh}.
One can interpret $\sqrt{\langle(\triangle H^*)^2\rangle}$ as a measure
for the deviation of the simulation from a reference case in
which only $W_-$ fluctuations are simulated.
This simulation was done with a chain-contour
discretization of $\dd s=0.01$, which allows for the calculation of
$\triangle H$ to within 0.01. Typically, in simulations involving the
$W_-$ field alone, we used $\dd s=0.05$ in order to save computing
time. The discretization error of $\triangle H$ in this case is
increased to $\lesssim 0.08$. Considering that typical values of
$\triangle H$ 
are on the order of $\pm 3$, this appears like a very 
\\

\vspace*{-1cm}

\fig{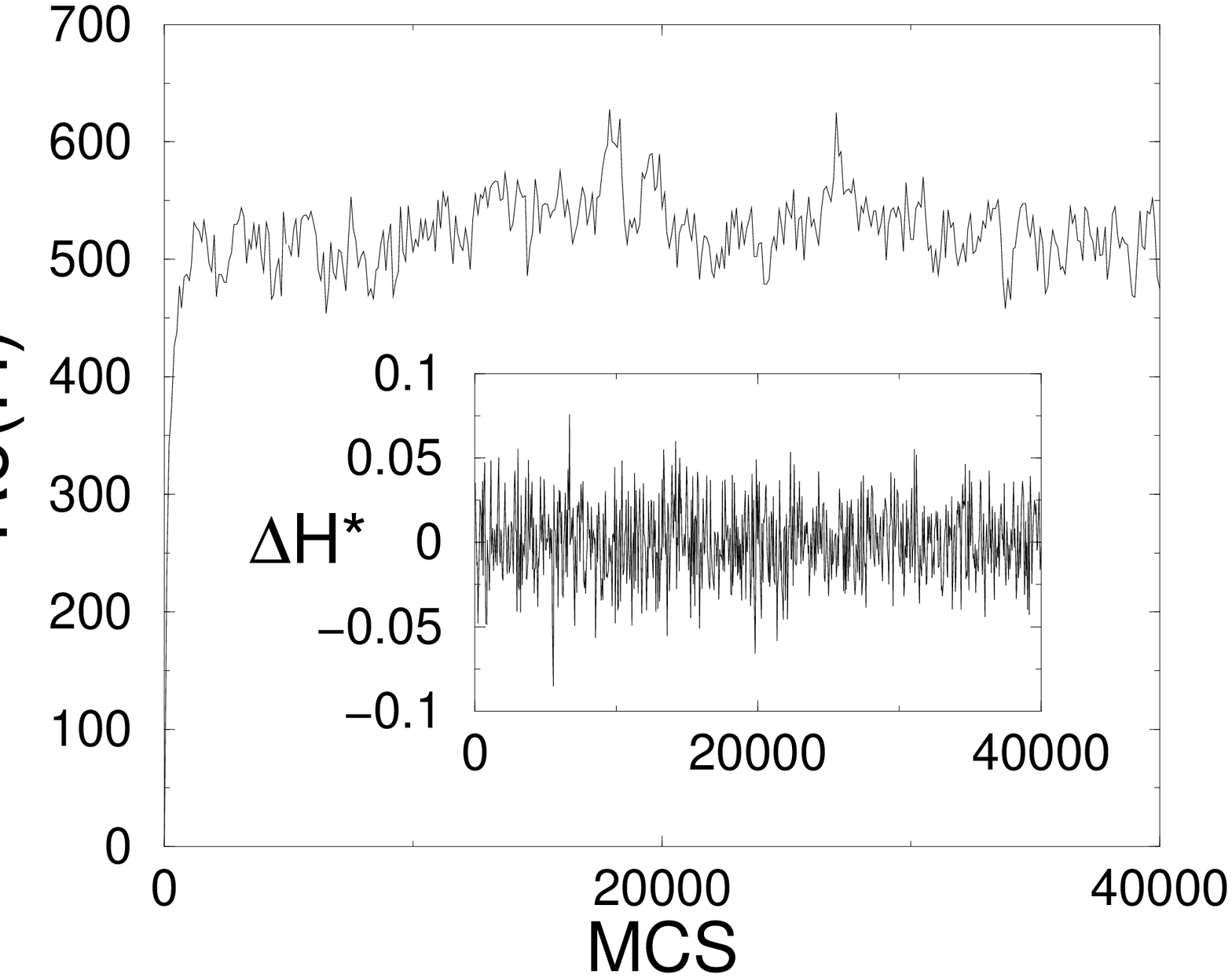}{67}{60}{}

\CChh

\noindent
moderate numerical error. At the same time, we realize that
the effect of the fluctuations in $W_+$ expressed through
$\triangle H^*$ is well below our standard numerical accuracy:
$\sqrt{\langle(\triangle H^*)^2\rangle} \approx 0.022$. Furthermore, the
average of $\triangle H^*$ very closely approximates zero, so
there is no drift of $\triangle H$ as a result. We therefore
conclude that contributions of the $W_+$ fluctuations to $H^R$
cannot be distinguished as such from numerical inaccuracies.

Next, we examine the reweighting factor, $I_{GC}$, in the particular
simulation discussed above. Numerically, we must evaluate:

\begin{equation}
\label{IGCR}
I_{GC}^R = \cos\left(C\left[\int \dd \br\tilde\omega_+ + \sum_i z_i
Q_i^I\right]\right)\equiv\cos(A)
\end{equation}

The two terms in the argument, $A$, of the cosine in (\ref{IGCR}) have
opposite signs and largely cancel out each other, as was demonstrated
theoretically above. However, in practice, the numerical application of
$I_{GC}$ to the simulation is impeded greatly by the fact that
the two contributions to $A$ are extensive, so the fluctuations in
$A$
are amplified. $I_{GC}$ is a non-positive-definite weighting
factor, echoing the well-known ``sign problem'' encountered in
fermionic field theories in elementary particle physics
\cite{signproblem}. Fig.~\ref{fig:awplus} displays part of a
time series of $A$ in our above test run. It performs fast oscillations
around a very small absolute mean value which within numerical accuracy
closely approximates zero when averaged over the
entire run, although it is expected from theory to be finite. This
\\

\vspace*{-1cm}

\fig{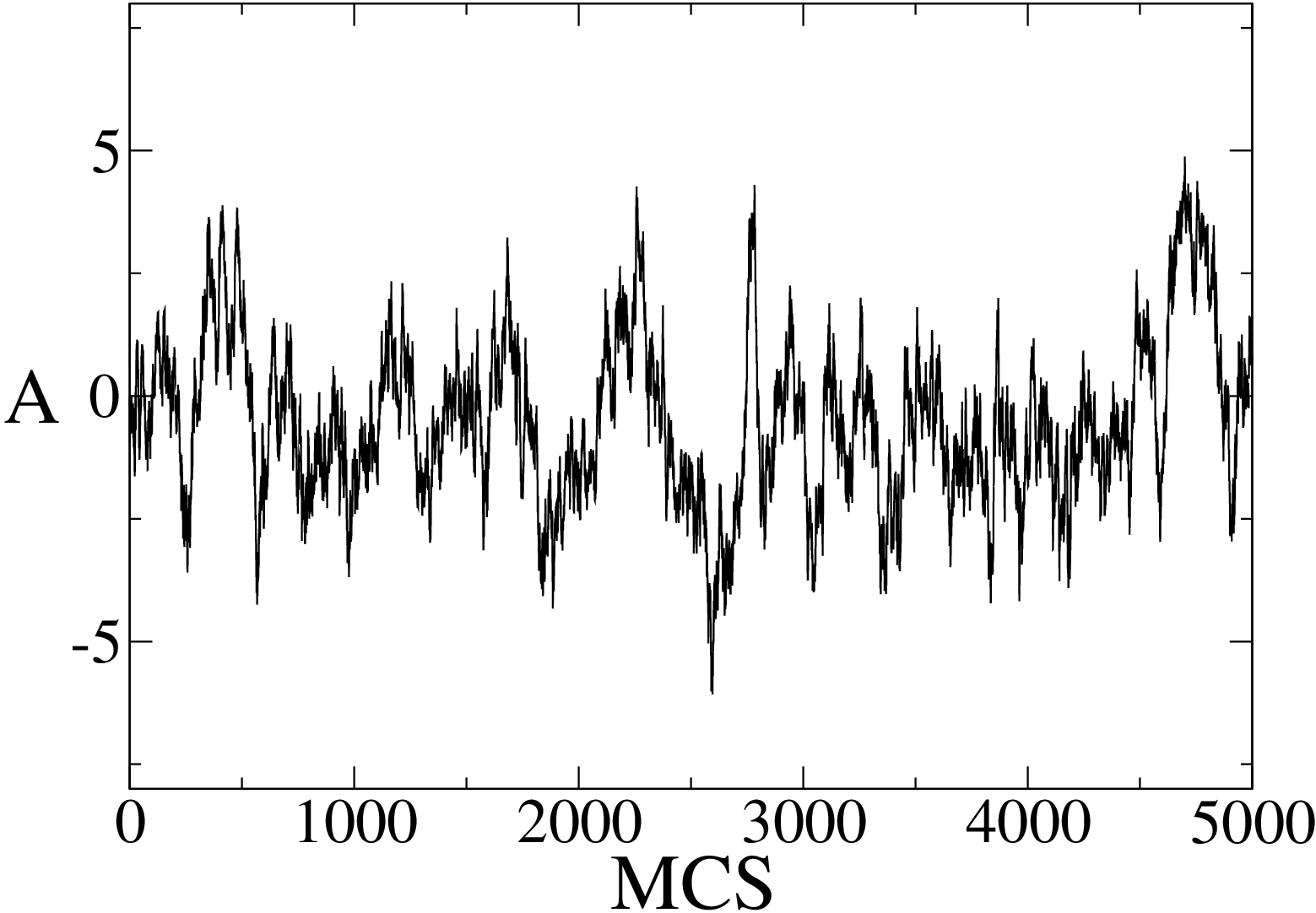}{70}{55}{}
\CCawplus

\noindent
is the essence of our sign problem: the statistical deviation from the 
mean completely dominates $A$. At this point, we seem to have arrived 
at a dead end, having to concede that we cannot numerically implement 
the reweighting factor. 

However, we still may get a sense of the impact
of the $\tilde\omega_+$ fluctuations from examining pairwise
correlation functions of $\tilde\omega_+$, $W_-$, and $I_{GC}$. The
correlation function is defined as
\begin{equation}
C[A,B;t] = \frac{\langle A(t')B(t'+t)\rangle - \langle A \rangle
\langle B \rangle}{\sigma_A^2 \sigma_B^2},
\end{equation}
where $\sigma$ denotes standard deviation, and $t$ and $t'$ are numbers of
moves in $W_-$. If A or B is a field, we average over all lattice points 
${\bf r}$. Fig.~\ref{fig:wmwm} shows $C[W_-,W_-]$, and in 
Fig.~\ref{fig:correl}, $C[I,I]$, $C[\tilde\omega_+,\tilde\omega_+]$, 
$C[I,W_-]$, and $C[\tilde\omega_+,W_-]$ are displayed. On comparing the 
various correlation times, an important qualitative difference is 
observed: while the autocorrelation time of $W_-$ is on the order of
$10^5$ Monte Carlo steps in $W_-$ (subsequently denoted MCS), those of
$\tilde\omega_+$ and $I$ are only on the order of a few hundred MCS,
i.e., several orders of magnitude smaller\cite{footnote2}.
Furthermore, the correlation functions of $W_-$ with both
$\tilde\omega_+$ and $I$ are virtually flat lines, indicating the
absence of any cross-correlations.
The same is obtained for $C[\tilde\omega_+,W_-^2]$.

With this result in mind, consider again the partition function
(\ref{Z_GC}). We demonstrated before that we may essentially replace
$H^R[W_-,\tilde\omega_+]$
with $H[W_-]$ and so rewrite $Z_{GC}$ as
\begin{equation}
Z_{GC} \propto \int \! \! {\cal D}W_- e^{-H_{GC}}
\int \! {\cal D} \tilde\omega_+ I_{GC}[W_-,\tilde\omega_+].
\end{equation}
Since, as we have seen, $I$ and $W_-$ are virtually uncorrelated, we argue
that the results of the simulation of $W_-$ cannot possibly be altered by
the reweighting factor in a significant way.

\fig{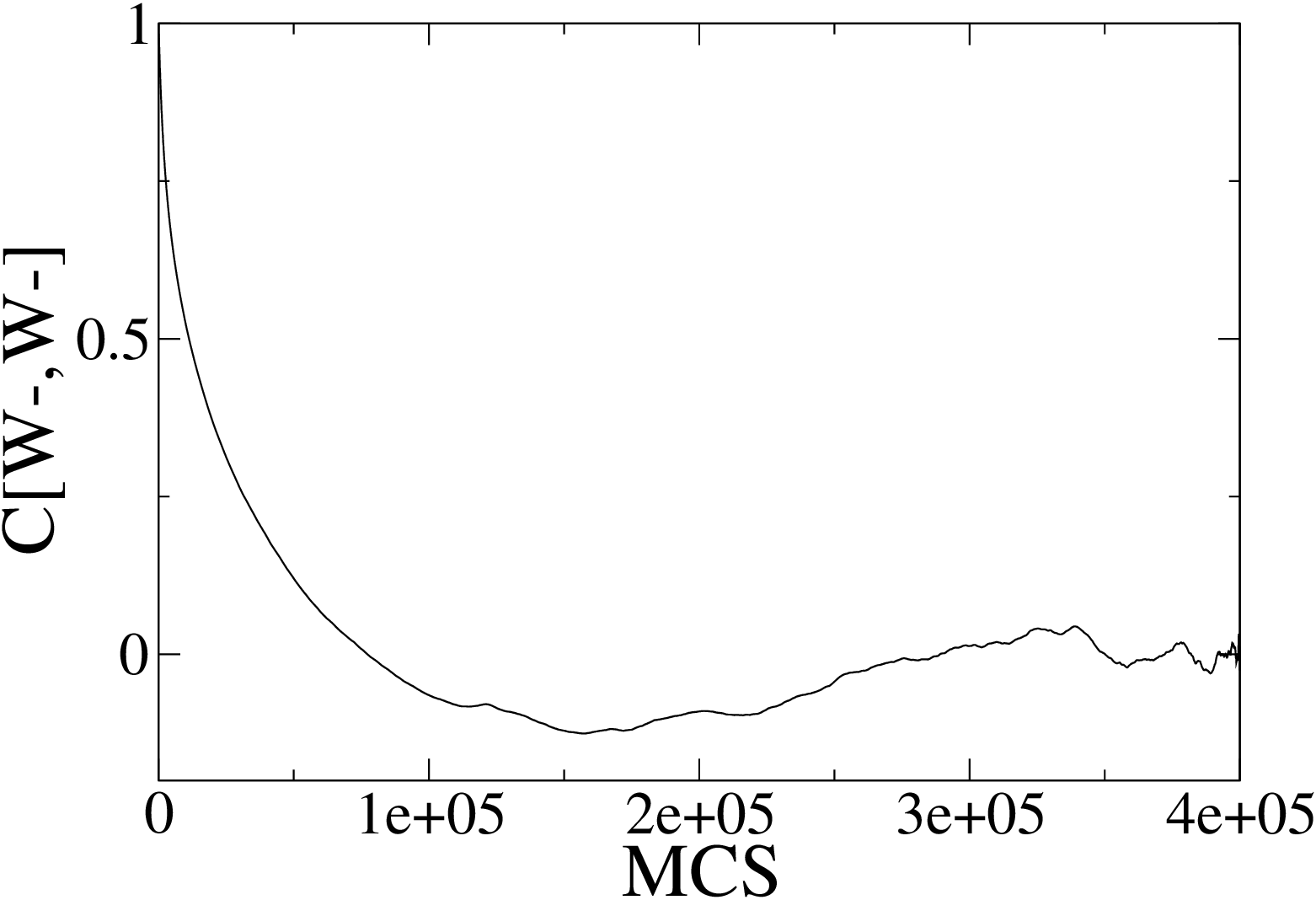}{68}{52}{}
\CCwmwm

\fig{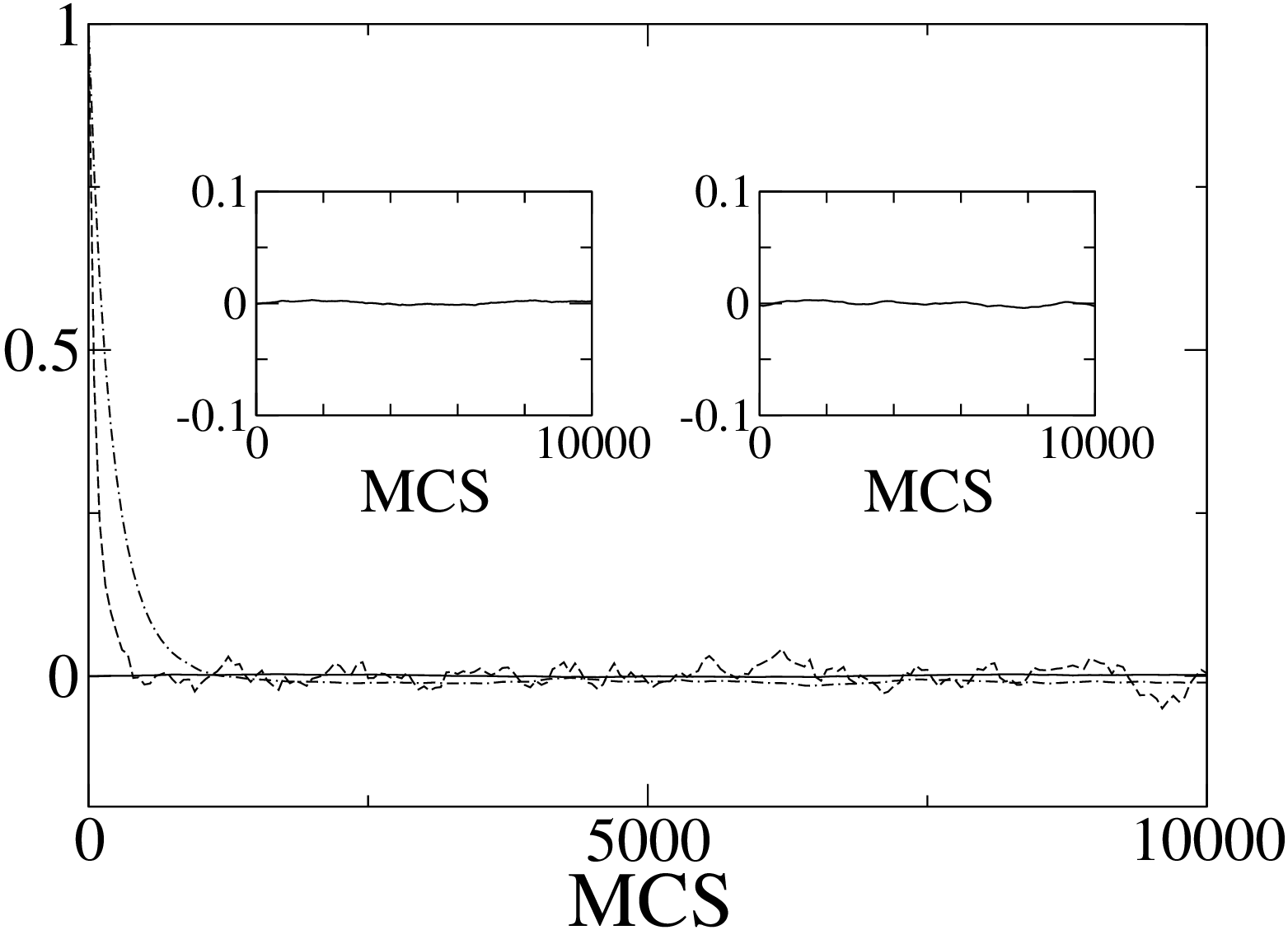}{65}{50}{}
\CCcorrel

In sum, the results of these test simulations indicate that the
$W_+$ fluctuations do not influence the structure of the polymer
blends substantially. Therefore, the Monte Carlo simulations presented
in the next section were restricted to $W_-$ fields only.

\section{Results and discussion}
\label{results}
We have studied the two-dimensional ternary A+B+AB system at 
dimensionless number density $C=50$. In evaluating these simulations, we
must establish two phase boundaries: first, the
fluctuation-shifted order-disorder transition, and second, the
onset of the phase-separated region. For the former, two
parameters have been used which we shall define in the following.

{\em The Direction Persistence parameter}: To define and determine
this parameter,
a given distribution $W_-(\br)$ is translated via (\ref{barphi})
into a ``density'' pattern $\bar\phi_A(\br)$. All simulations
performed with the field-theoretic Monte Carlo method that are
presented in this section took only $W_-$ fluctuations into
account. Hence, $\bar\phi_A(\br)$ is real. Lattice points are
denoted black if $\bar\phi_A(\br)<0.5$, and white if
$\bar\phi_A(\br)\ge 0.5$. We further define $\mbox{$\langle
\max(l_{\bf a}) \rangle $}$ as the maximal length of either black
or white sections along a one-dimensional cross section of the
image in direction ${\bf a}$, averaged over all offsets along the
$X$ axis (or equivalently the $Y$ axis). The direction persistence
parameter, $\Lambda$, is defined for a two-dimensional
black-and-white image as
\begin{equation}
\Lambda :=
\left<\frac{\langle \max(l_\parallel) \rangle}
{\langle \max(l_\perp)\rangle}\right>_{^{\Big |}\mbox{all
directions}} - 1,
\end{equation}
where $l_\parallel$ denotes the direction being averaged over, and
$l_\perp$ the direction perpendicular to a given $l_\parallel$.
$\Lambda$ is 0 in a disordered configuration and
positive definite in a lamellar configuration, and thus
constitutes a measure for the ``lamellarness'' of a configuration.
Furthermore, it is dimensionless and does not directly depend on
the lattice size. $\Lambda$ measures lamellarness from a local
perspective.

{\em Anisotropy Parameters}: As a complementary measure of anisotropy, we
define the parameters $\bar{F}_2$ and $\bar{F}_4$, which measure the
anisotropy of the Fourier transform,
$F({\bf q})$, of a $\bar\phi_A (\br )$ image to look at
periodicity from a more macroscopic point of view:
\begin{equation}
\label{Sn}
F_n(q) := \frac{1}{2\pi} \Big| \int_0^{2\pi} \!\!\!
\dd \phi |F({\bf q})|^2 e^{i n\phi} \Big|.
\end{equation}
By definition, $F_2$ and $F_4$ are identically zero in a
disordered (isotropic) configuration and are positive otherwise. A
more convenient way to analyze $F_n(q)$ is to calculate a normalized
ratio of the
integral and the standard deviation of $F_n(q)$, which yields a
single dimensionless number:
\begin{equation}
\bar{F}_n := \frac{\int \! \dd q F_n(q)}{\sigma(q)|_{F_n}},
\end{equation}
\begin{equation}
\sigma(q)|_{F_n} \equiv
\Big[ \frac{\int \! \dd q \: q^2 F_n(q)}{\int \! \dd q \: F_n(q)}
- \Big(\frac{\int \! \dd q \: q \: F_n(q)}
{\int \! \dd q \: F_n(q)}\Big)^2
\Big]^{1/2}.
\end{equation}
$\bar F_n$, like $F_n(q)$, is 0 for disordered and nonzero for
lamellar configurations. Note that
$F({\bf q})$ is not the structure factor of the melt but simply
the Fourier transform of $\bar\phi_A$. In fully equilibrated
simulations, the structure factor can be calculated from averaged
expressions containing the fields $W_-$ and $W_+$ \cite{glenn}.
\\

\vspace*{-1.cm}

\hspace*{-0.6cm}
\fig{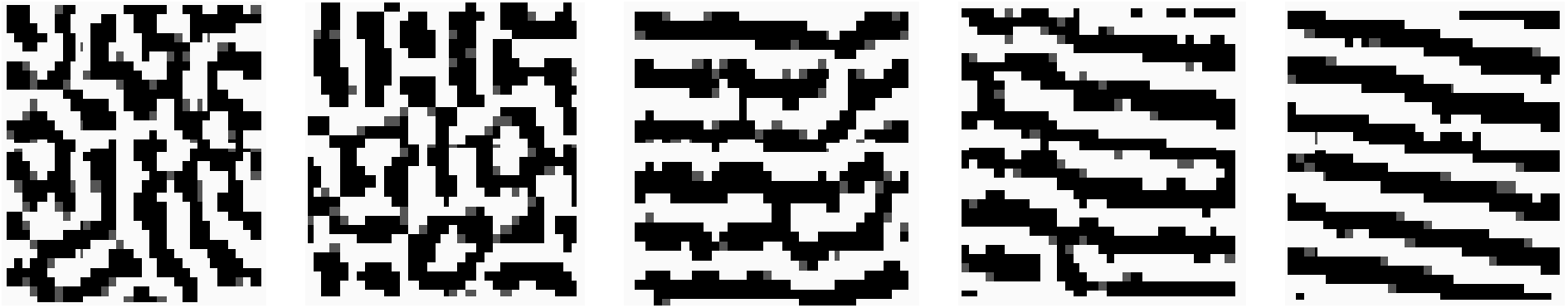}{70}{25}{}
\CCsnapmc

\subsection{Monte Carlo simulations}

We shall first discuss the results of the Monte Carlo simulations.

The shift of the order-disorder transition was examined in the
canonical ensemble on a $32 \times 32$ lattice with $\dd x$=0.625 for
$\phi_H=0$ and $0.2$, and $\dd x$=0.78125 for $\phi_H=0.4$, $0.6$, and $0.7$.
These $\dd x$ values were chosen such that a single lamella was several
pixels wide, which is a precondition for the evaluation of the $\Lambda$
parameter to work properly. For simplicity, we shall subsequently
denote runs that started from a disordered configuration
``D-started'' and those that started out of a lamellar
configuration, ``L-started.'' In the latter, the lamellar
periodicities of the starting configurations reflect the
mean-field values at the given parameters. Moreover, in L-started
runs, $\dd x$ was chosen to allow for an integer number of lamellae
to fit in the simulation box.

In the case of copolymers only ($\phi_H = 0$), D-started runs at
various $\chi N$ yield the snapshots of Fig.~\ref{fig:snapmc}, which
were taken after approximately 1 million Monte Carlo steps and
after the simulations were equilibrated. It is clearly visible
from the images that above a certain threshold $\chi N$, a
lamellar phase begins to form spontaneously. We did not observe
any hysteresis effects. We should stress that the patterns
$\bar\phi_A(\br)$ shown on these snapshots are not to be confused
with real density patterns -- they are merely visualizations of
the field distributions $W_-(\br)$.

To quantify the transition, we have
plotted $\Lambda$ in Fig. \ref{fig:lambda} and $\bar{F}_2/ \bar{F}_4$ in
Fig. \ref{fig:f2f4}. These plots also include curves for the other runs
described above, i.e., homopolymer volume fractions $\phi_H$ up to $0.7$.
In all cases, a jump is displayed in both the $\Lambda$ and
$\bar{F}_2$/$\bar{F}_4$
parameters at the fluctuation-corrected critical $\chi N$. For $\phi_H=0$,
we find the transition to be shifted from the mean-field value of
10.495 to 11.3(1), which is in good agreement with a result
obtained earlier by two of us with the complex Langevin 
method \cite{venkat}. It is also further evidence that
the $W_+$ fluctuations contribute only a minor correction to the
partition function.

For homopolymer concentrations $\phi_H$ above 0.7 the configurations
did not spontaneously assemble into a lamellar phase for any $\chi N$
in D-started simulations on $32 \times 32$ lattices.
The configurations can at best be described as very defective
lamellae. However, in L-started runs the configurations only broke
up at very low $\chi N$ values. For example, at $\phi_H=0.85$,
D-started runs were done for $\chi N \le 13$, all of which stayed
disordered, yet in L-started runs the lamellae only broke up for
$\chi N < 12.1$. This obvious discrepancy can be made plausible by
two arguments: First, in the D-started runs, as we approach higher
$\chi N$ values (which correspond to lower temperatures), we
observe a freezing effect, i.e., the simulations get trapped in
configurations with a certain degree of {\it local} periodicity
yet the systems cannot move through configuration space
efficiently enough to achieve quasi-long-range 

\vspace{0.5cm}

\fig{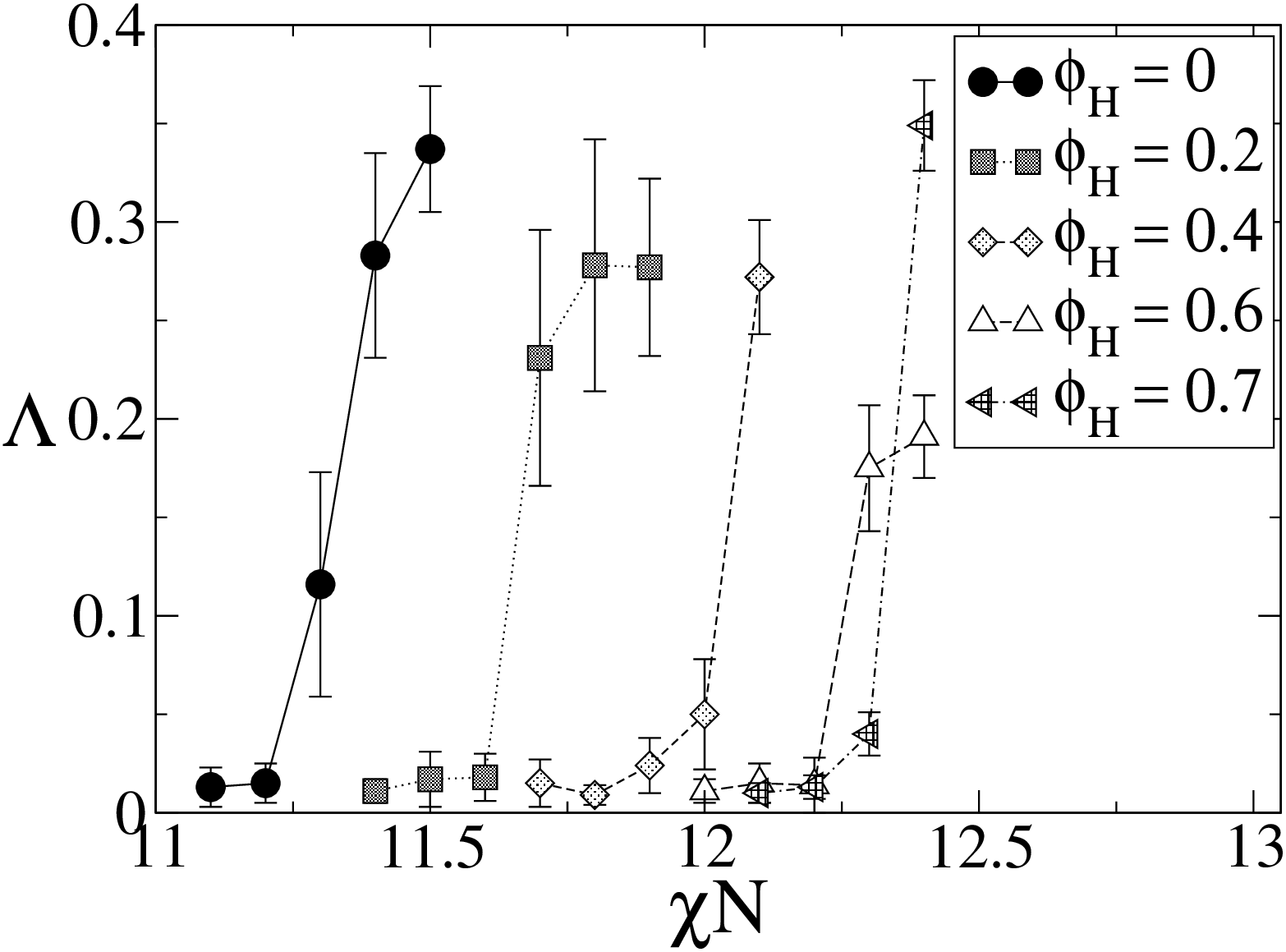}{60}{45}{}
\CClambda

\fig{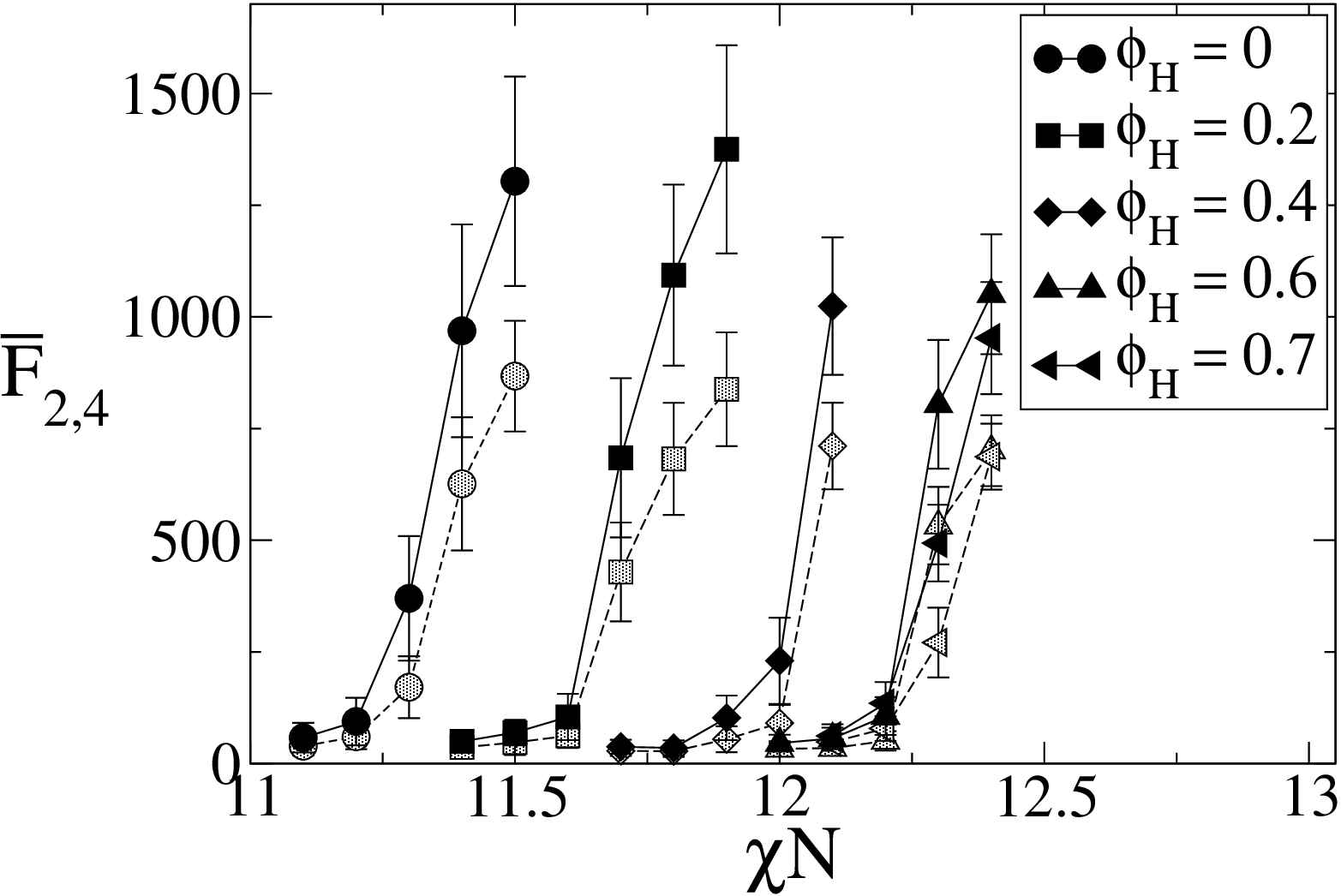}{65}{45}{}
\CCff

\bigskip

\vspace*{3.3cm}

\fig{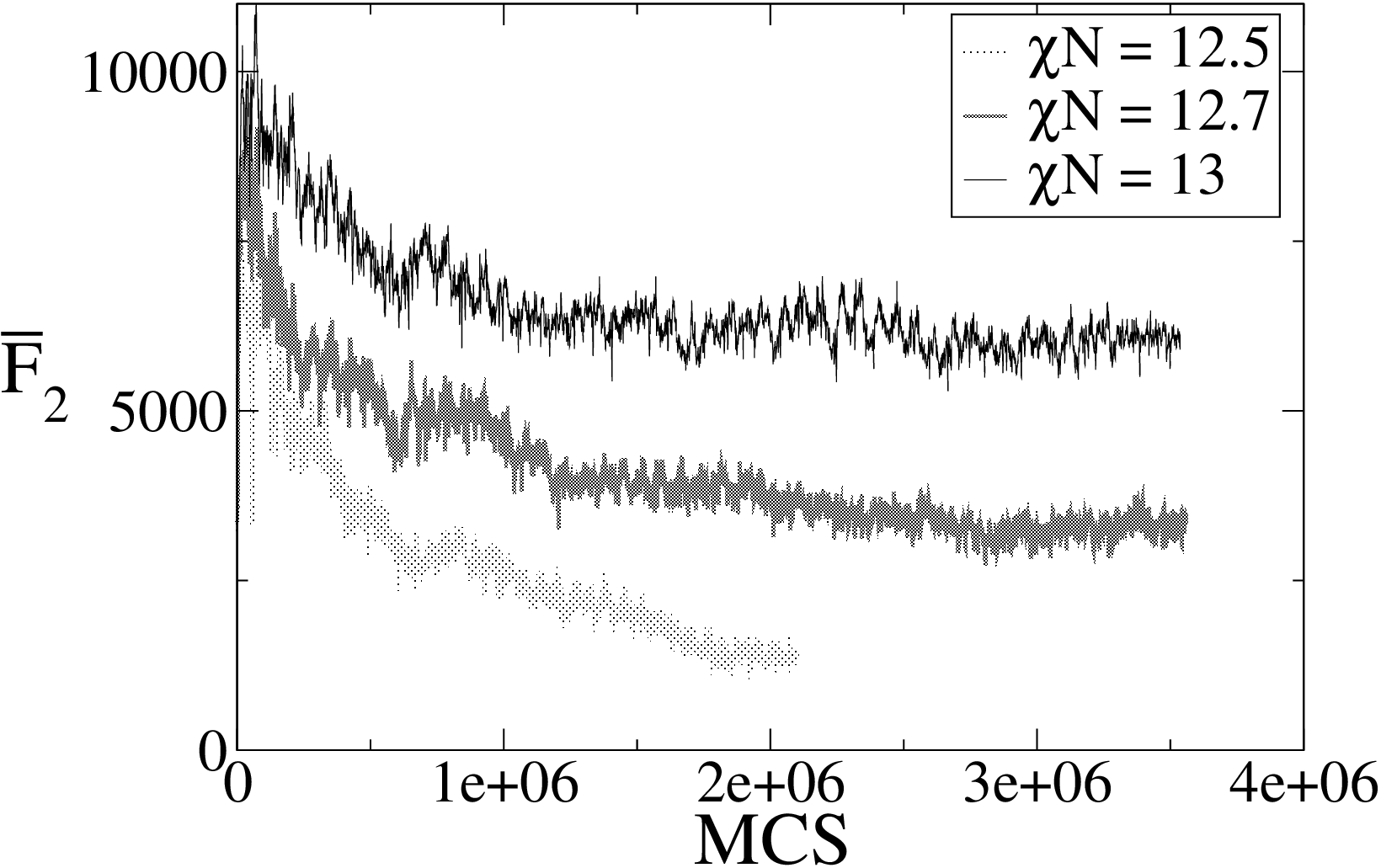}{65}{50}{(b)}

\vspace*{-9.7cm}

\fig{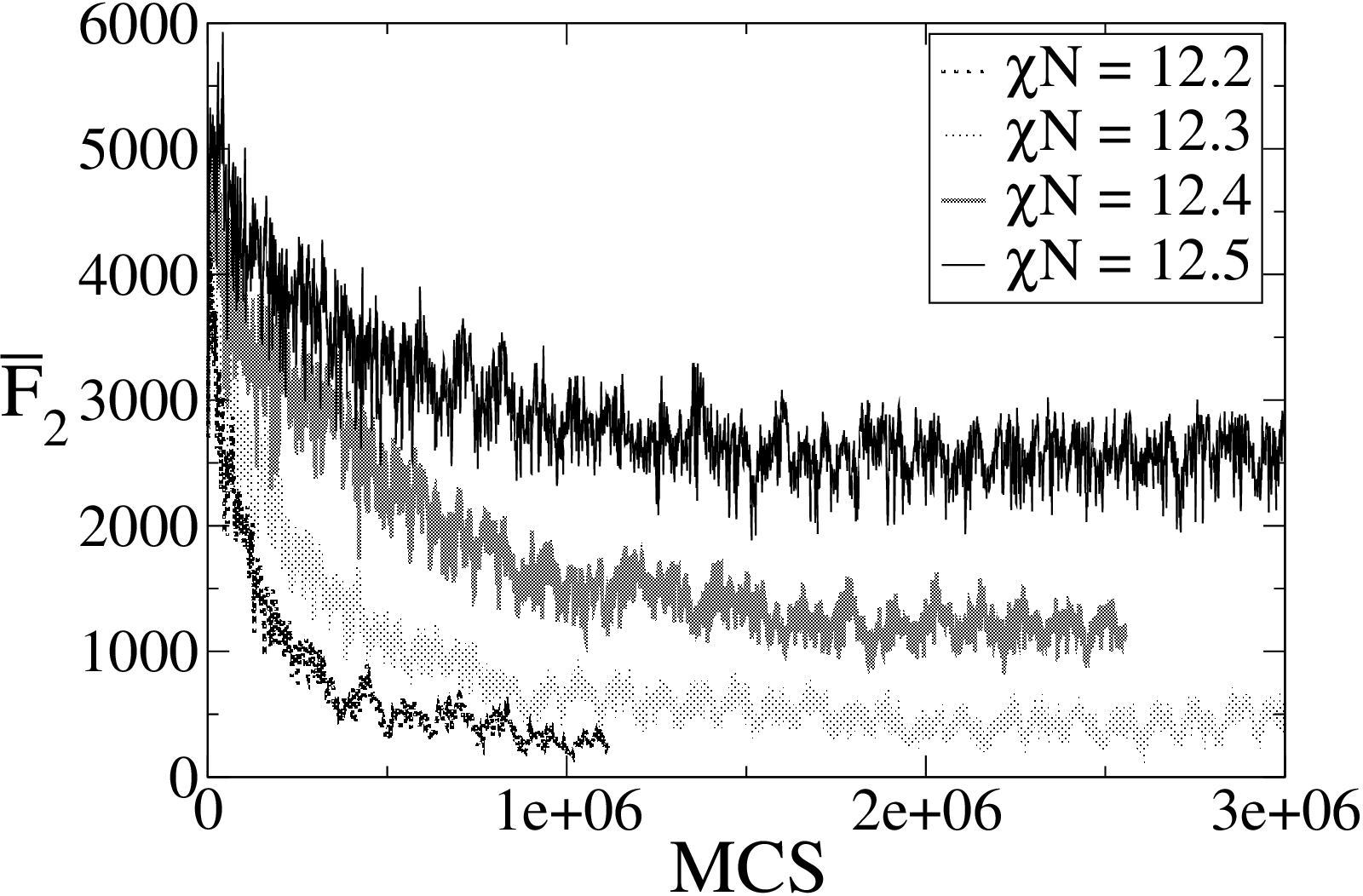}{65}{50}{(a)}

\vspace*{4.2cm}

\CCfseries

\noindent
order. Second, in
the L-started runs, in order to keep $\dd x$ reasonably low (i.e.,
$\lesssim 0.8$), we can only put a small number of lamellae in the
box. In the case of $\phi_H=0.85$, that number is 3. But this in
combination with the periodic boundary conditions imposed on the
box in turn artificially stabilizes the lamellar phase, which is
why we do not see a breaking up. Evidently, we must use bigger
simulation boxes for $\phi_H\gtrsim 0.7$.

Because of the freezing effects at high $\chi N$ described above,
we restrict ourselves to L-started runs in examining the
order-disorder transition on $48 \times 48$ lattices. Even so, we find that
the simulations take an increasingly greater number of Monte Carlo
steps to equilibrate the higher $\chi N$. For $\phi_H=0.7$, a time
series of $\bar{F}_2$ is displayed in Fig. \ref{fig:fseries} (a).
The plot indicates that the transition must be between
$\chi N$ = 12.3 and 12.5: we see a substantial jump in
both parameters, caused by a breaking-up of the initial lamellar
configuration. This is in good agreement with the result from the
$32 \times 32$ lattice. The $\bar{F}_2$ graphs at the transition,
especially for $\chi N$ = 12.3, display a special feature:
a periodic oscillatory motion once the run is equilibrated. The time
scale of this oscillation can be explained as follows. 
Close to the transition, the lamellae do not easily dissolve completely.
Instead, defects develop which may open and close periodically,
often alternating the continuation of a given contour length between
two defective lamellar branches. This oscillation is reflected in
$\bar{F}_2$.

For $\phi_H=0.8$, the corresponding time series is displayed in
Fig. \ref{fig:fseries} (b). At $\chi N=12.5$, the system is clearly
disordered, and the transition is inferred to occur below $\chi N=13$.
Another run for $\phi_H=0.85$ indicates that that transition is at
approximately $\chi N=13$, although even after 3 million Monte Carlo steps
the configurations retained a strong correlation to the seed
configuration. We conclude that on the one hand, using a bigger lattice has
alleviated the artificial stabilizing effect of the periodic boundary
conditions on the lamellae. On the other hand, however, we still encounter
numerical freezing effects for $\chi N \gtrsim 13$ at C=50, which
prohibit the examination of this region of the phase diagram.

As concerns the order of the ODT, we did sweeps in the grand
canonical ensemble across the ODT for relative homopolymer fugacities
$z=1$ and $z=2$ (Fig. \ref{fig:conc}). Both plots are consistent 
with a very weakly first-order transition. 
In either case, the coexistence
region has a width of no more than approximately 0.001 in $\Phi_H$.

The onset of the (macro) phase-separated two-phase region was examined in 
the grand canonical ensemble; a D-started run at parameters in the
phase-separated region will eventually undergo

\vspace*{4.5cm}

\fig{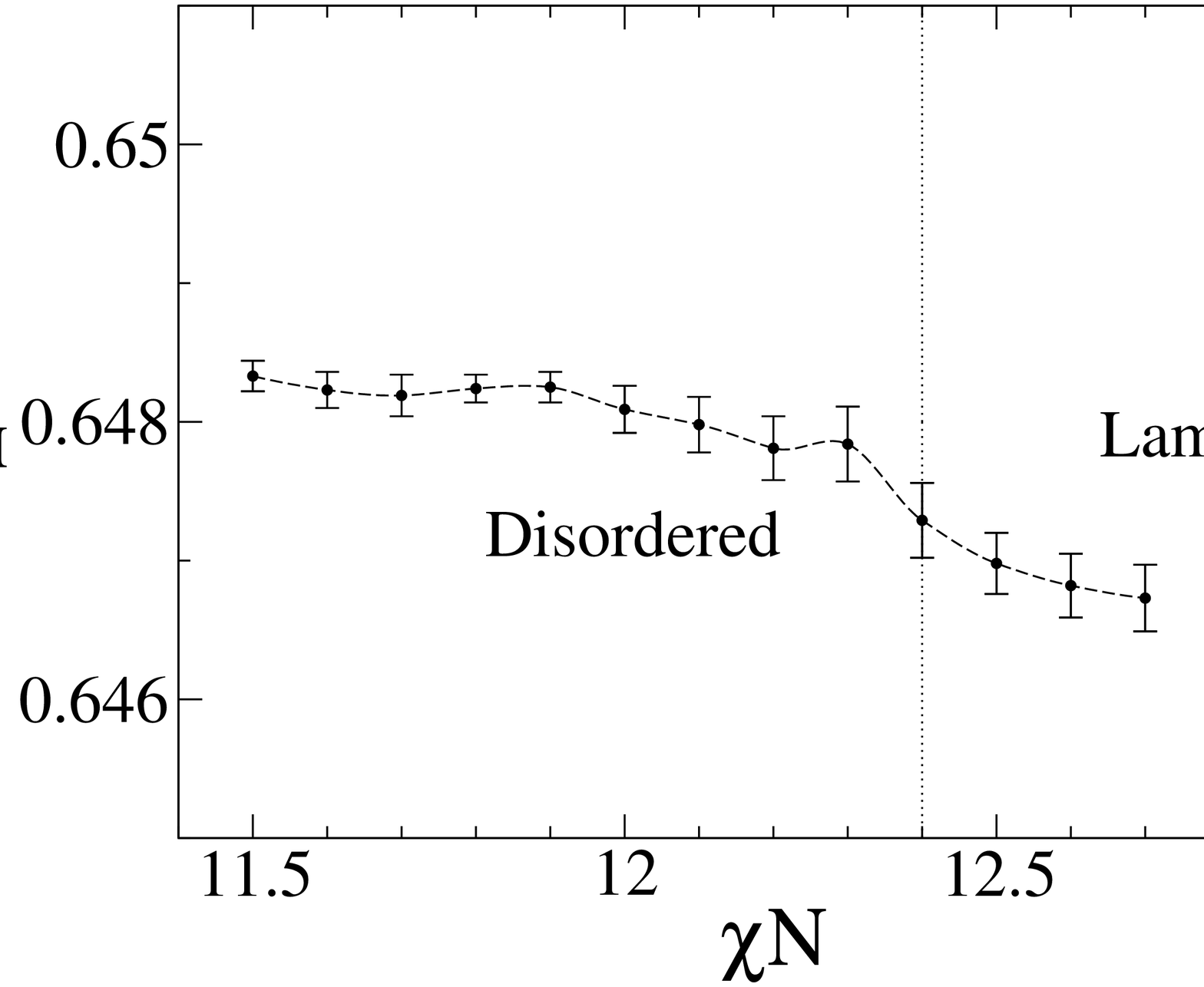}{65}{49}{(b)}

\vspace*{-9.3cm}

\fig{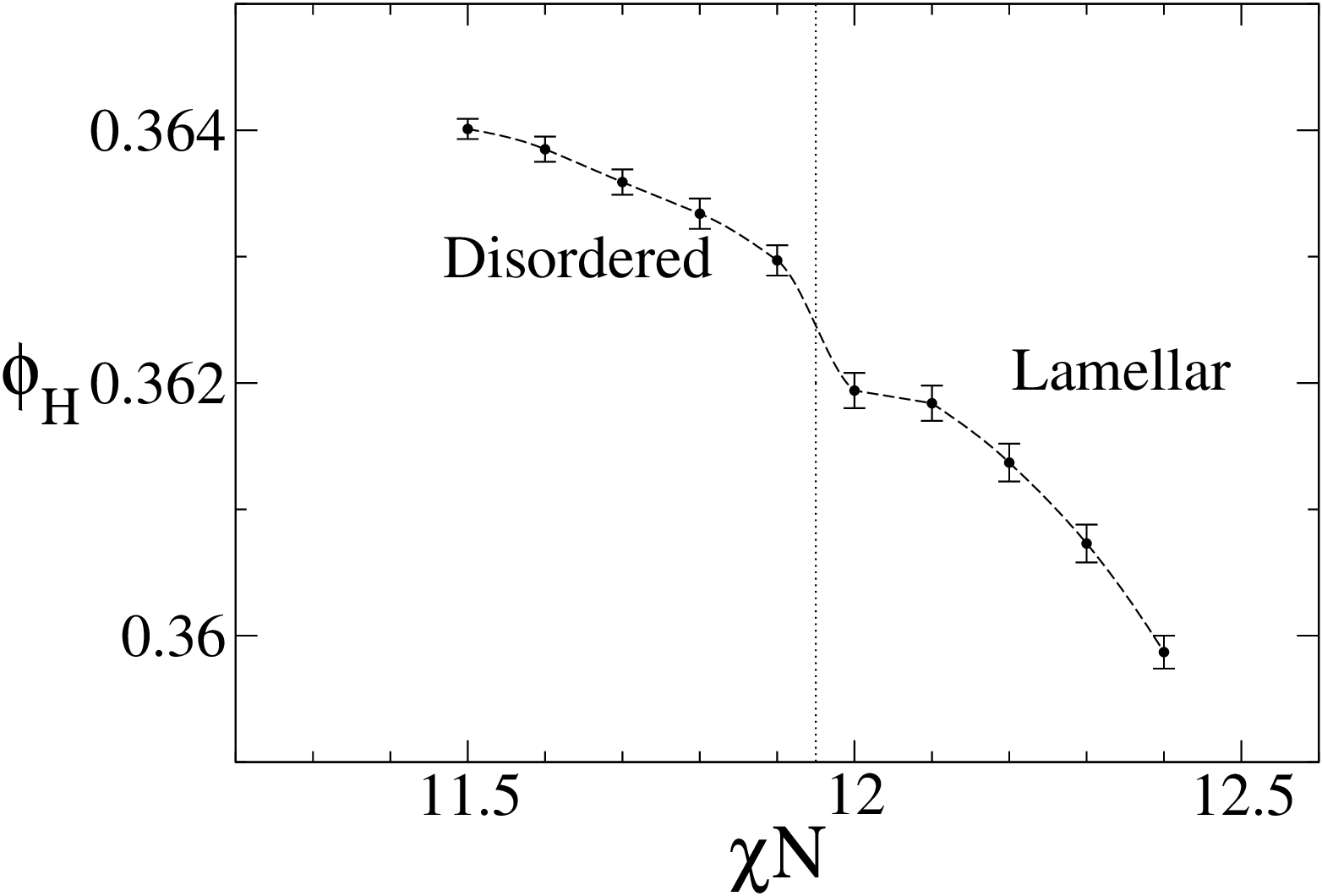}{63}{45}{(a)}

\vspace*{4.2cm}

\CCconc

\noindent

\fig{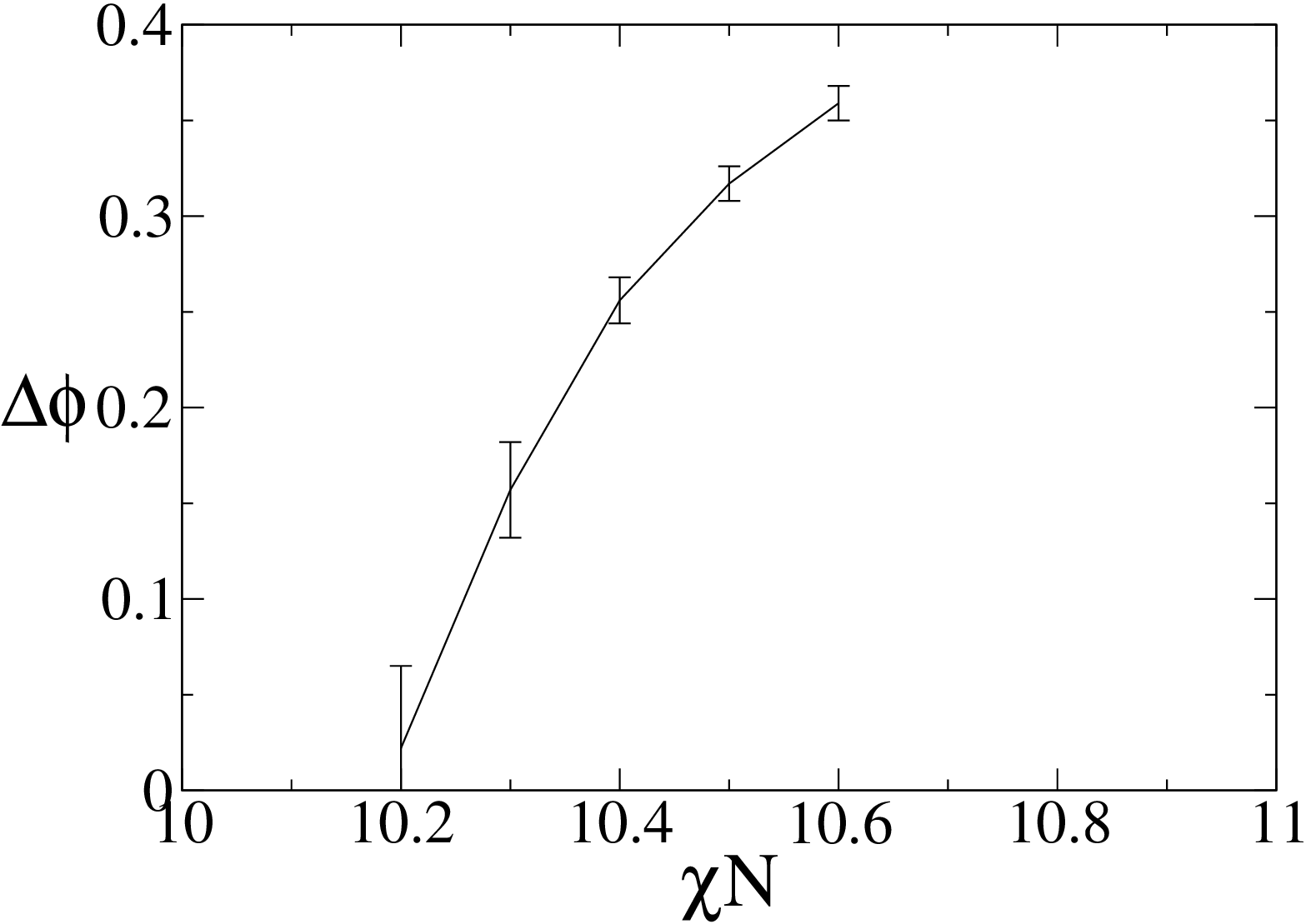}{65}{47}{}
\CCabdemix

\fig{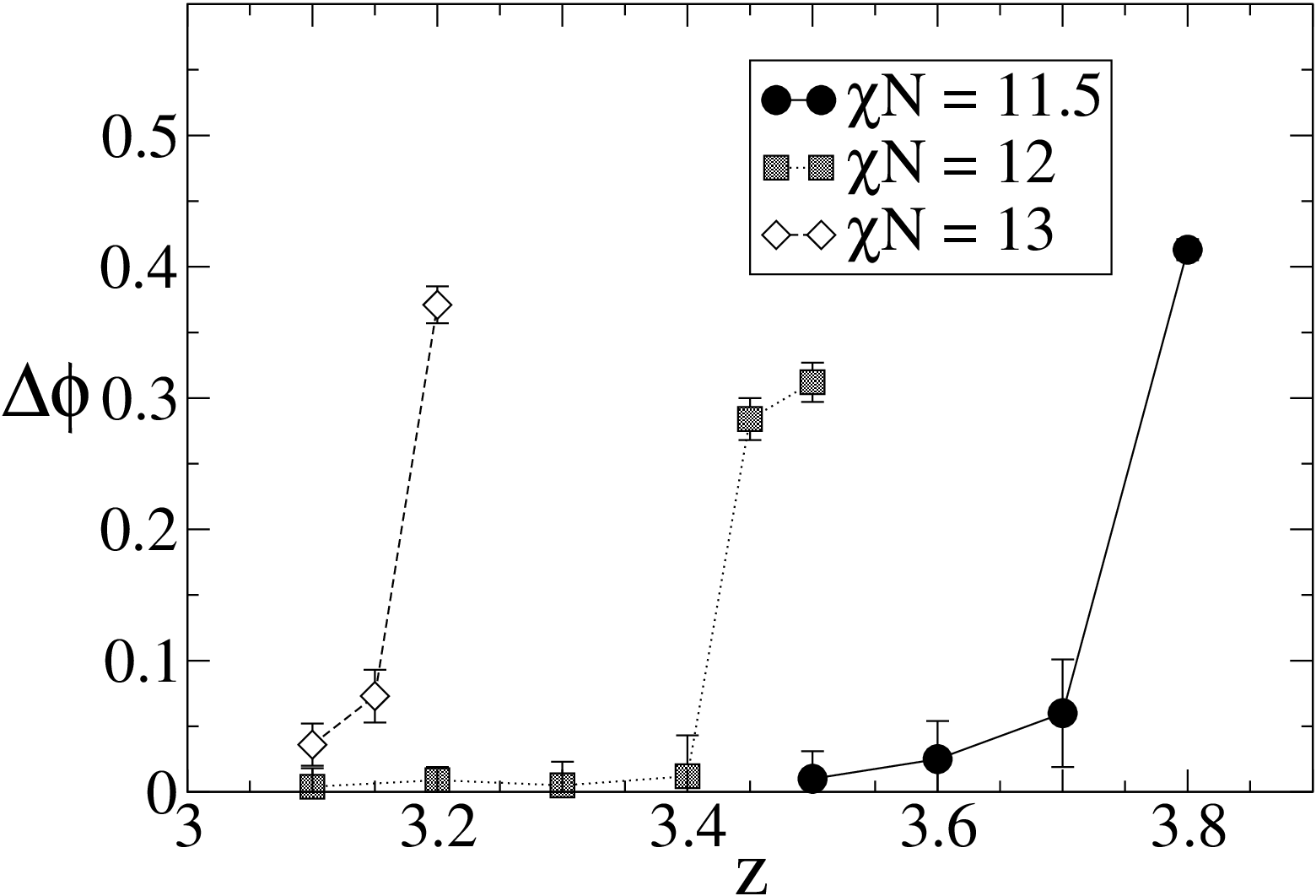}{65}{47}{}
\CCpc

\noindent
a spontaneous symmetry
breaking and end up in a configuration dominated by either A or B
segments. Moreover, it is sufficient to do these runs on $32 \times 32$
lattices as no complex morphologies are investigated here. To quantify
this behavior, we analyze the total difference of normalized A and B
densities:
\begin{equation}
\triangle\phi := \left| V^{-1} \int \dd {\bf r} (\bar{\phi}_A({\bf
r}) - \bar{\phi}_B({\bf r})) \right |.
\end{equation}

First, we looked at the demixing in a system consisting only of A and B
homopolymers ($\phi_H=1$). In this case, Fig. \ref{fig:abdemix} shows how 
the order parameter $\triangle\phi$ grows from zero above approximately
$2.04/\alpha$,
shifted up from the mean-field value of $2/\alpha$. For various $\chi N$,
Fig. \ref{fig:pc} displays $\triangle\phi$ for relative homopolymer
fugacities $z$
(corresponding to $\phi_H<1$ in the canonical ensemble) around the 
transition.  When translated into $\phi_H$ values in the canonical 
ensemble, the numerical accuracy obtained in the present work
did not allow for determining the order of the corrected 
disorder/two-phase
transition.
\\

\vspace*{-1cm}

\fig{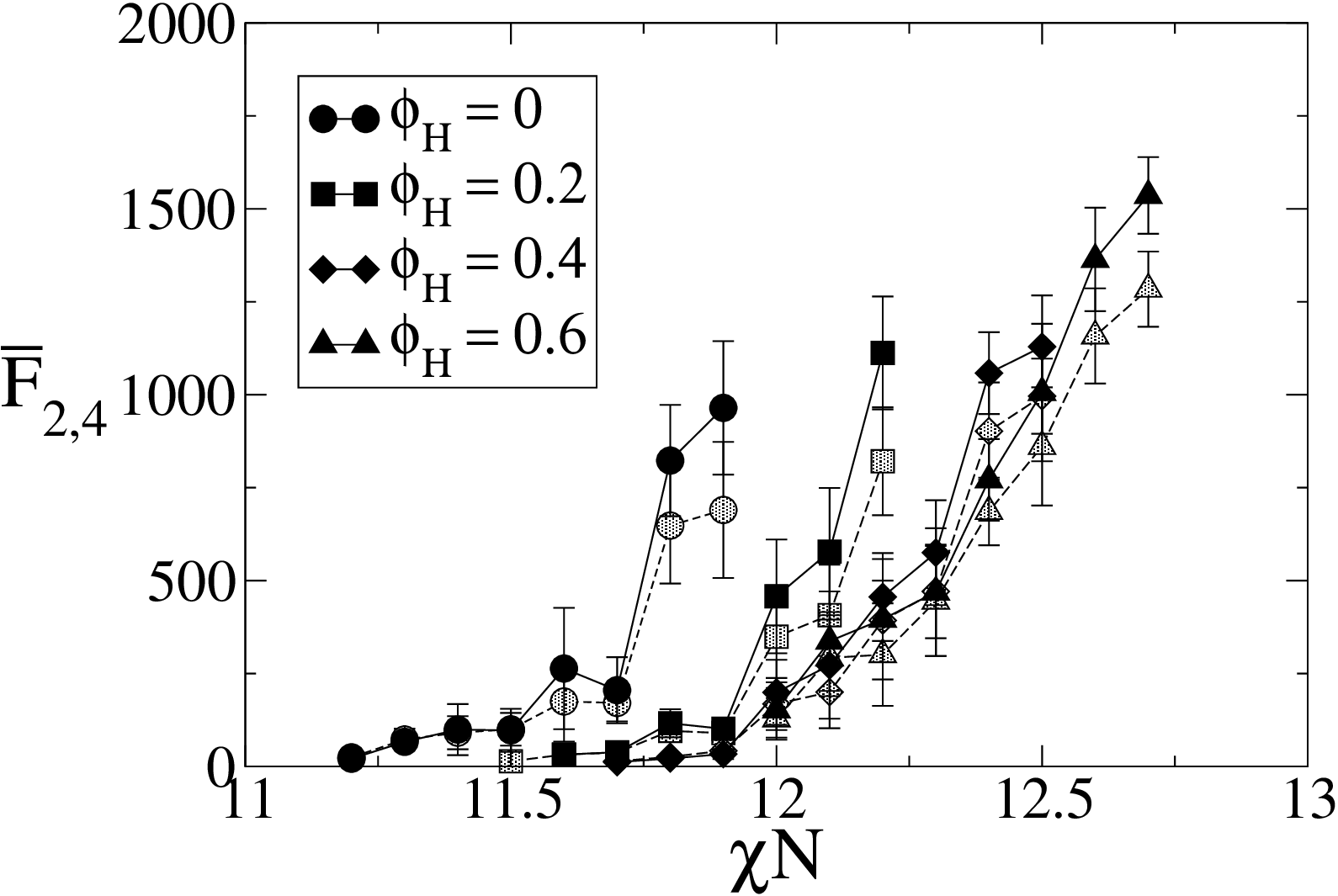}{67}{52}{}
\CCvgf

\fig{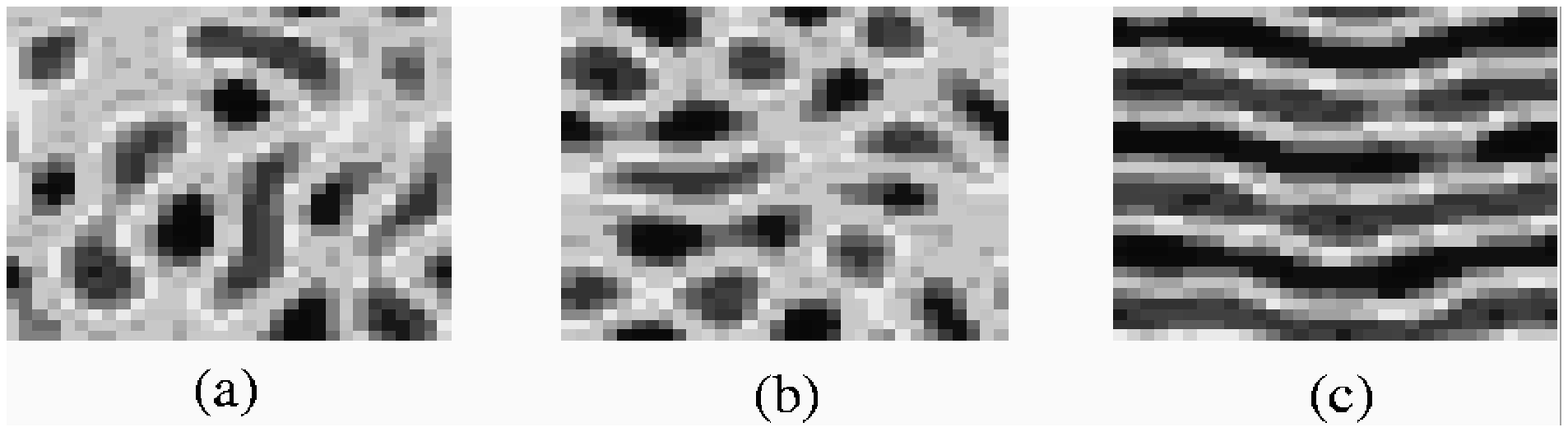}{65}{20}{}
\CCsnapvg

\subsection{Complex Langevin simulations}

The same system has also been examined by complex Langevin
simulations for selected parameter values in the canonical ensemble.
The simulations were initiated from L-started configurations
(in the lamellar phase) to examine the shift in the ODT.
The transition points were identified by examining the values
of the anisotropy parameters $\bar{F}_2$ and $\bar{F}_4$. The behavior of
these quantities (displayed in Fig.~\ref{fig:vgf}) is quite similar
to that observed in the field-theoretic Monte Carlo approach,
and exhibits a steep increase near the onset of microphase
separation. As an additional pictorial proof of such a
transition, Fig.~\ref{fig:snapvg} displays the averaged values of the
{\em real} component of the fluctuating density fields across the
transition.

As mentioned earlier in the text, complex Langevin
simulations were hindered by numerical constraints arising from
extensive phase fluctuations near the Lifshitz point, and required
small time steps to maintain numerical stability. However, the use of
such time steps 

\noindent
led to freezing effects, and rendered equilibration
difficult. Consequently, we have displayed the results from CL
simulations only for situations (i.e.,~$\phi_H$ values) for which the
fluctuation-corrected transition temperatures could be identified
reliably. It is encouraging to note that the results obtained from
the CL simulations, the basis of which is somewhat distinct from the
field-theoretic Monte Carlo method presented in this article, do
indicate very good agreement with the results of the latter approach.
The implications of such results are threefold:
(i) This serves to corroborate the results presented in earlier
sections which were obtained using the Monte Carlo method.
(ii) It also suggests that the fluctuation effects in the
$W_+$ field, which were neglected in the Monte Carlo simulations
(but which were included in the CL approach), have only a negligible
influence upon the majority of the phase diagram.
(iii) Further, the coincidence of CL results with the
Monte Carlo approach suggests that despite the tenuous foundations
of complex Langevin approach, such a method does yield physically
reliable results for a major portion of the phase diagram,
and thereby warrants further studies using such an approach.

\subsection{The fluctuation-corrected phase diagram
and the formation of the microemulsion}

The results of our simulation investigations are put together in the
fluctuation-corrected phase diagram for C=50 (Fig. \ref{fig:ph_fluc}),
and lead to a slope of the resulting fluctuation-corrected
order-disorder transition which is in good qualitative agreement with
the experimental result of Hillmyer et al. for a PEE/PDMS/PEE-PDMS
system of intermediate molecular weights \cite{hillmyer}. In
particular, the shape of the cusp-like region of bicontinuous
microemulsion identified therein is here reproduced
computationally for the first time.
%
It is important to note that to minimize computational expense our
simulations were carried out in two dimensions, while the
experiments are inherently three-dimensional. In general, one
would expect the fluctuation effects to be stronger in 2-d than in
3-d and, indeed, the microemulsion ``cusp'' appears broader in
Fig. \ref{fig:ph_fluc} than in the experimental phase diagrams.
Nevertheless, we believe that a firm qualitative correspondence
between simulation and experiment has been established.

\hspace*{-0.7cm}
\fig{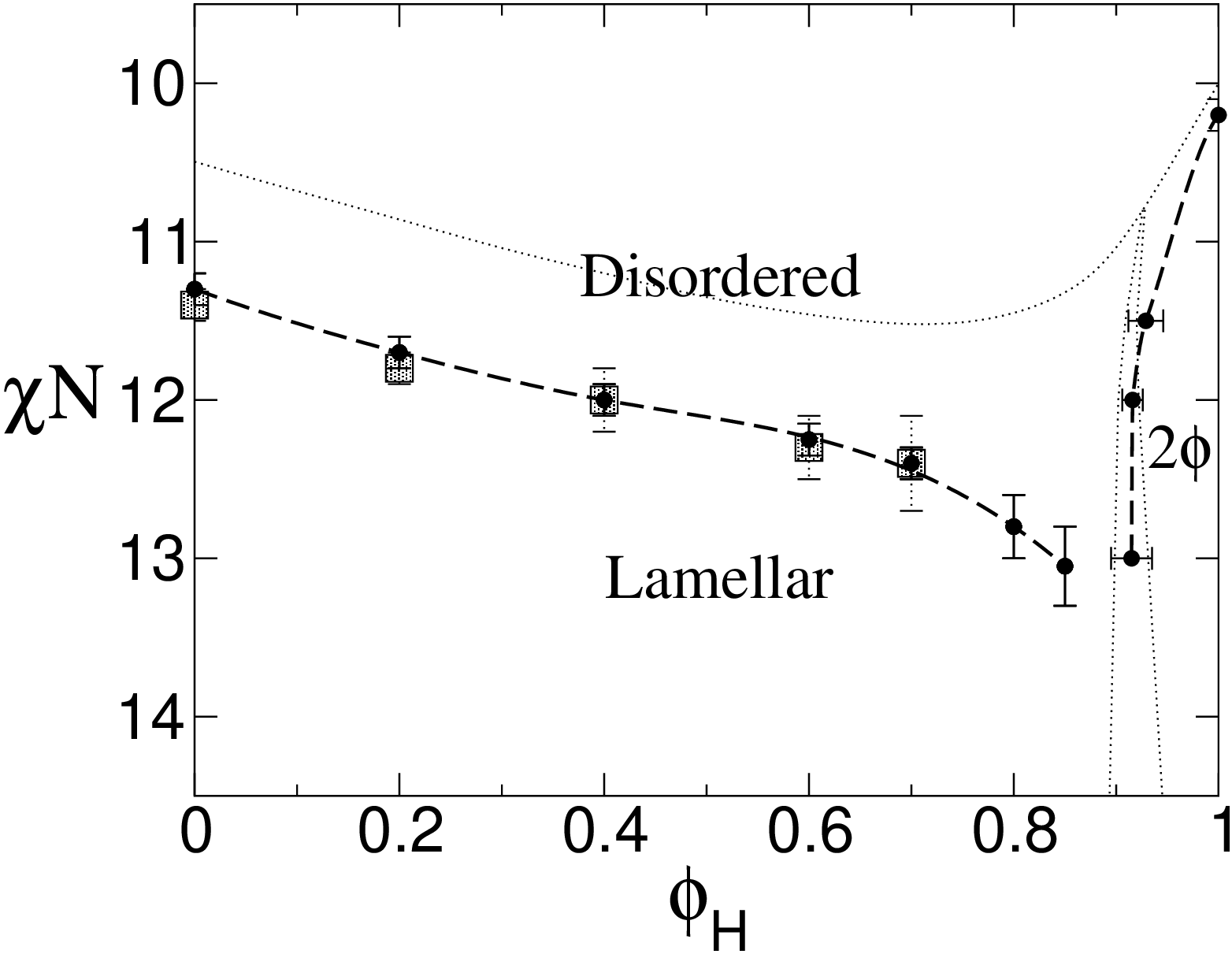}{70}{57}{}
\CCphfluc

\fig{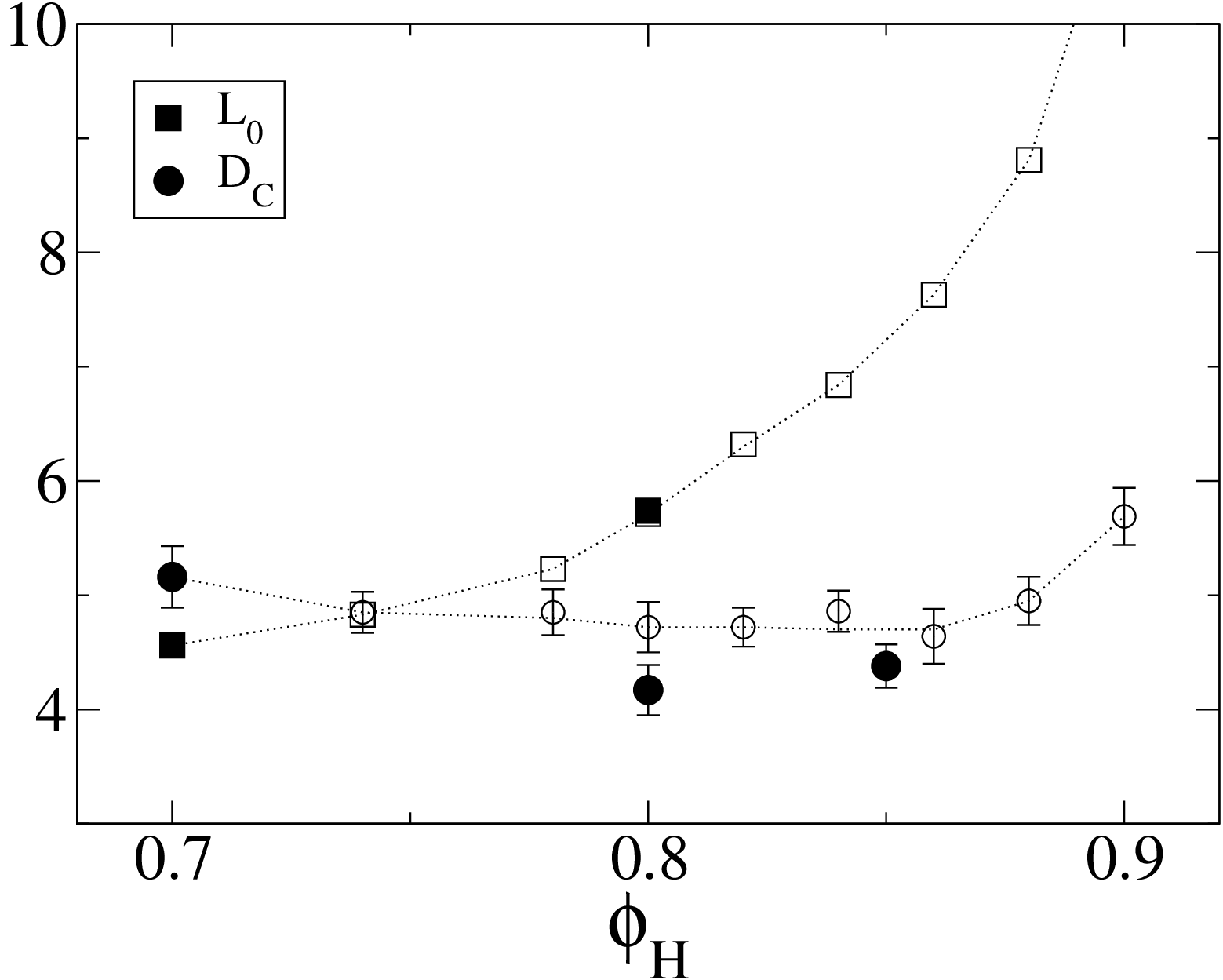}{60}{50}{}
\CClength

To understand better what happens at the onset of the microemulsion
phase, we have calculated the mean curvature diameter, $D_C$, of the
boundaries of A and B microdomains:
\begin{equation}
D_C = 2 \left[\frac{1}{L_c}\int \dd s
\left|\frac{d{\bf t}}{ds}\right|^2\right]^{-1/2},
\end{equation}
where $L_c$ is the sum of all contour lengths of the microdomain
boundaries, and {\bf t} is the tangent vector at a given
coordinate $s$ along the contour. Details of the calculation
of this object will be presented elsewhere \cite{future}.
Furthermore, we examine the maximum, $q_{max}$, of $\bar{F}_0(q)$,
which indicates the preferential Fourier wave number of a
configuration multiplied by $2\pi$. In turn, $L/q_{max}=:L_0$
indicates the preferential length scale of a configuration. Fig.
\ref{fig:length} compares $D_C$ and $L_0$ for various homopolymer
concentrations $\phi_H$ at $\chi N = 12.5$, for both L-started and
D-started runs. Whereas for lower $\phi_H$, $D_C$ is 
larger than
$L_0$, the two lengths cross each other as $\phi_H$ is increased.
Thus we have identified the mechanism by which fluctuations
generate the microemulsion phase: it forms when the width of the
lamellae grows larger than the curvature radius of the fluctuating
boundaries, causing the lamellae to break up.
This type of ``Lindemann criterion'' for production of a
bicontinuous microemulsion by melting of a lamellar phase has
discussed elsewhere in the context of oil/water/surfactant phases
\cite{dcmorse}.

\section{Summary and conclusions}
\label{summary}
In the present paper, we have introduced a new way of evaluating
the full partition function of a system of polymeric blends
subject to thermal fluctuations. The complex Langevin approach which
was employed in our earlier researches, accounts for both the
$W_-$ and $W_+$ fluctuations, and is able to predict the shift
in the order-disorder transition at low $\phi_H$ and in the two-phase
region. For $0.7 \lesssim \phi_H \lesssim 0.9$ it runs into numerical
difficulties.  We believe that the large system sizes
necessitated in such regions as well as the concomitant
strong fluctuation effects arising near the Lifshitz points
restricts the applicability of the complex Langevin method in such
regimes. As an alternative, we have proposed  a field-theoretic
Monte Carlo approach, which  was shown to be applicable to greater
regions of the phase diagram. Small
 lattices ($32 \times 32$) sufficed up to $\phi_H=0.7$, above which bigger
($48 \times 48$) lattices had to be used. In the region where both
methods work, they are in good agreement, indicating that the
$W_+$ fluctuations neglected in the latter simulations contribute
only a minor correction. Further evidence for this circumstance
was found in empirical test runs. Such observations also suggest
that one could have potentially employed a real Langevin scheme
where the $W_+$ fluctuations are neglected \cite{reister}.
Regretfully, however, a full analysis of the $W_+$ fluctuations
was rendered difficult by the occurrence of a
non-positive-definite weighting factor (the sign problem). A
fluctuation-corrected phase diagram was presented for C=50, which
corresponds well in qualitative terms to the PEE/PDMS/PEE-PDMS
system of intermediate polymerization indices examined in a series
of recent experiments \cite{bates1,bates2,morkved,hillmyer},
providing further corroboration of the existence of a region of
bicontinuous microemulsion between the lamellar and
phase-separated regions. The mechanism responsible for the
formation of this phase was shown to be intimately related to
matching length scales of the curvature of microdomain boundaries
and of the lamellar periodicity. In an upcoming article, we will
examine properties of the disordered phase \cite{future}.

We have benefited from fruitful discussions with M. Schick, R.
Netz, M. Matsen, and S. Sides. This work was supported by the
Deutsche Forschungsgemeinschaft (Germany). VG
and GHF were partially supported by the National Science
Foundation under Awards DMR-02-04199 and DMR-98-70785, and the
Petroleum Research Fund, administered by the ACS. The Monte Carlo
simulations were carried out on the CRAY T3E of the NIC institute
in J\"ulich.

\section*{Appendix: Complex Langevin Method}
In this appendix, we provide a brief outline of the complex
Langevin method. Readers interested in more details are referred to
Ref.~\cite{glenn}. The complex Langevin technique involves
generating $W_-$ and $W_+$ states {\em in the entire complex
plane} (despite the fact that the integrations in Eq.~(\ref{eq8})
are restricted to the real axis in $W_-$ and the imaginary axis in
$W_+$) by means of {\em complex Langevin} equations. Specifically,
the equations used to generate the real and imaginary parts of the
$W_-$ and $W_+$ fields, $W_-^R, W_-^I, W_+^R,$ and $W_+^I$, are of
the form:
\begin{eqnarray}
\frac{\partial [W_-^R (\br ,t) + i W_-^I (\br ,t)]}{\partial t} & = &
-\frac{\delta H [W_-, W_+]}{\delta W_-(\br ,t )}
\nonumber \\ &&
 + \: \theta (\br ,t),
 \label{eq16a}
 \end{eqnarray}

 \begin{eqnarray}
\frac{\partial [W_+^I (\br ,t) - i W_+^R (\br ,t)]}{\partial t} & = &
-i\frac{\delta H [W_-, W_+]}{\delta W_+(\br ,t )}
\nonumber \\ &&
+ \theta (\br ,t),
\label{eq16}
\end{eqnarray}
where $\theta (\br ,t)$ is a {\em real} Gaussian white noise with
moments satisfying the appropriate fluctuation-dissipation
relation: $\langle \theta \rangle =0$,  $\langle \theta (\br ,t)
\theta (\br^\prime ,t^\prime ) \rangle = 2 \delta (t-t^\prime )
\delta ( \br - \br^\prime )$. In the above equations,
\begin{eqnarray}
\frac{\delta H[W_-,W_+]}{\delta W_- (\br )} &= &
C \: {\bigg [} \frac{2 W_-}{\chi N} - {\bar{\phi}}_A (\br ; W_- , W_+)
\nonumber \\ &&
+ \: {\bar{\phi}}_B (\br ; W_-, W_+ ) {\bigg ]}
\label{eq17}
\\
\frac{\delta H[W_-,W_+]}{\delta W_+ (\br )} &=& C \:
[{\bar{\phi}}_A (\br ; W_- , W_+ ) \nonumber \\ && + \:
{\bar{\phi}}_B (\br ; W_- , W_+ ) - 1] \label{eq18}
\end{eqnarray}
The evolution of the $W_-^R, W_-^I$ and $W_+^R, W_+^I$ fields
according to the above Langevin equations
(\ref{eq16a} - \ref{eq16}) produces in the long-time limit the
generation of field states such that the ``time'' averages acquire
the same values as those computed using the original
non-positive-definite Boltzmann weight \cite{parisi}.
Equilibration of the simulation can be established by monitoring
an appropriate quantity, such as the average of the imaginary
component of energy/densities which should identically vanish at
equilibrium. We note that with such an averaging procedure,
physical observables such as the average A monomer density,
$\phi_A = \langle \bar{\phi}_A \rangle$, turn out to be real,
while their fluctuating analogs such as $\bar{\phi}_A$ are
generally complex.

It is pertinent to note a few special features of the complex
Langevin equations (\ref{eq16a} - \ref{eq16}). In the absence of
the noise term $\theta (\br ,t)$, the dynamics
will drive the system to the nearest saddle point. Consequently,
it is straightforward to probe the
stability of the mean-field solutions within the CL simulation
scheme by setting the noise term identically to zero. Furthermore,
this feature enables the sampling in the presence of the noise to
be automatically confined near a saddle point,
avoiding a long equilibration period if the system were to be
prepared in an initial configuration far from the saddle point.


\begin{thebibliography}{99}
\bibitem{footnote} Incompressibility here means that the number
  of monomers per volume in a given volume is constant, 
  regardless of the size of the volume. This is possible for 
  Gaussian chains, because they are scale invariant on small 
  scales.
\bibitem{footnote2} The values of the correlation times depend on the 
  details of the simulation algorithm. In simulation runs with a 
  different ratio of $\tilde{\omega}_+$ and $W_{-}$ moves, 
  the autocorrelation times of $W_-$, $I$ and $\tilde{\omega}_+$ 
  may become comparable. Nevertheless, our conclusion that the
  time scale of $\tilde{\omega}_+$ fluctuations is not coupled to 
  that of the $W_{-}$ fluctuations should remain valid, since all
  simulation schemes are equivalent.
\end{thebibliography}
\end{document}